\documentclass[useAMS,usenatbib]{mn2e}

\usepackage{graphicx}
\usepackage{amssymb}

\title[ULAS~J222711$-$004547]{The extremely red L dwarf ULAS~J222711$-$004547 - dominated by dust.}
\author[F. Marocco et al.]{F. Marocco$^{1}$\thanks{E-mail:
f.marocco@herts.ac.uk;}, A. C. Day-Jones$^{1}$, P. W. Lucas$^{1}$, H. R. A. Jones$^{1}$, R. L. Smart$^{2}$, \newauthor Z. H. Zhang$^{1}$, J. I. Gomes$^{1}$, B. Burningham$^{1}$, D. J. Pinfield$^{1}$, R. Raddi$^{3}$, L. Smith$^{1}$ \\
$^{1}$Centre for Astrophysics Research, Science and Technology Research Institute, University of Hertfordshire, Hatfield AL10 9AB, UK. \\
$^{2}$INAF/Osservatorio Astrofisico di Torino, Strada Osservatorio 20, 10025 Pino Torinese, Italy. \\
$^{3}$Department of Physics, University of Warwick, Coventry CV4 7AL, UK.
}

\begin{document}

\date{Accepted XXXX. Received XXXX; in original form XXXX}

\pagerange{\pageref{firstpage}--\pageref{lastpage}} \pubyear{2013}

\maketitle

\label{firstpage}

\begin{abstract}
We report the discovery of a peculiar L dwarf from the UKIDSS Large Area Survey (LAS), ULAS~J222711$-$004547. The very red infrared photometry (MKO $J-K$ = 2.79$\pm$0.06, \textit{WISE W}1$-$\textit{W}2 = 0.65$\pm$0.05) of ULAS~J222711$-$004547 makes it one of the reddest brown dwarfs discovered so far. We obtained a moderate resolution spectrum of this target using VLT/XSHOOTER, and classify it as L7pec, confirming its very red nature. Comparison to theoretical models suggests that the object could be a low-gravity L dwarf with a solar or higher than solar metallicity. Nonetheless, the match of such fits to the spectral energy distribution is rather poor and this and other less red peculiar red L dwarfs pose new challenges for the modeling of ultracool atmospheres, especially to the understanding of the effects of condensates and their sensitivity to gravity and metallicity. We determined the proper motion of ULAS~J222711$-$004547 using the data available in the literature, and we find that its kinematics do not suggest membership of any of the known young associations. We show that applying a simple de-reddening curve to its spectrum allows it to resemble the spectra of the L7 spectroscopic standards without any spectral features that distinguish it as low metallicity or low gravity. Given the negligible interstellar reddening of the field containing our target, we conclude that the reddening of the spectrum is mostly due to an excess of dust in the photosphere of the target. De-reddening the spectrum using extinction curves for different dust species gives surprisingly good results and suggests a characteristic grain size of $\sim$0.5 $\mu$m. We show that by increasing the optical depth, the same extinction curves allow the spectrum of ULAS~J222711$-$004547 to resemble the spectra of unusually blue L dwarfs and even slightly metal-poor L dwarfs. Grains of similar size also yield very good fits when de-reddening other unusually red L dwarfs in the L5 to L7.5 range. These results suggest that the diversity in near infrared colours and spectra seen in late-L dwarfs could be due to differences in the optical thickness of the dust cloud deck.
\end{abstract}


\section{Introduction}
Brown dwarfs are sub-stellar objects, whose mass is insufficient to trigger and sustain stable thermonuclear fusion of hydrogen in their cores. Because they lack an internal source of energy brown dwarfs cool down over time, evolving through the spectral sequence from M to L, T and Y \citep[and references therein]{2012ApJ...753..156K,2005ARA&A..43..195K}. As a result, there is no unique mass $-$ spectral type relation for sub-stellar objects. For example, an L spectral class bin might be populated by an old low-mass star, a young high mass brown dwarfs, or even a very young planetary mass object.

Spectroscopy can provide useful insights to break this age-mass-luminosity degeneracy. Young brown dwarfs have in fact low surface gravity, as they have not contracted to their final radii, and some spectral features have been found to be sensitive to variations in the surface gravity. In particular, L dwarfs in young clusters show weak CaH, K I, and Na I absorption, and strong VO bands in their optical spectra \citep[and references therein]{2009AJ....137.3345C}. Near-infrared spectra show peaked $H$-band and a general flux excess towards longer wavelength \citep[e.g.][]{2001MNRAS.326..695L,2008MNRAS.383.1385L}. As a result, young brown dwarfs appear very red in terms of infrared colours.

Recent kinematic studies however have suggested that field objects showing signs of youth in their spectra could instead be relatively old \citep{2010ApJS..190..100K,2012ApJ...752...56F}. The peculiar morphology of their spectra therefore is not an effect of low surface gravity, but could instead be caused by an excess of dust in their photosphere \citep{2008ApJ...678.1372C} due to an higher-than-average metallicity \citep[e.g.][]{2007ApJ...655.1079L,2008ApJ...686..528L,2009ApJ...702..154S}. The nature of these objects, generally referred to as ``unusually red L dwarfs'' \citep[hereafter URLs, e.g.][]{2012AJ....144...94G} is not fully understood yet, in part because of the lack of such objects discovered so far. There are indeed only nine identified URLs: the L1pec 2MASS~J13313310+3407583 \citep{2008AJ....136.1290R,2010ApJS..190..100K}; the L5pec 2MASS~J18212815+1414010 \citep{2008ApJ...686..528L} and 2MASS~J23512200+3010540 \citep{2010ApJS..190..100K}; the L6.5pec 2MASS~J21481628+4003593 \citep{2008ApJ...686..528L} and 2MASS~J23174712$-$4838501 \citep{2008AJ....136.1290R,2010ApJS..190..100K}; the L7.5pec WISEP~J004701.06+680352.1 \citep{2012AJ....144...94G}; the recently discovered L7.5 WISE~J104915.57$-$531906.1A \citep{2013ApJ...767L...1L,2013ApJ...772..129B}; finally the L9pec WISEPA~J020625.26+264023.6 and WISEPA~J164715.59+563208.2 \citep{2011ApJS..197...19K}. 

Planetary-mass objects show similar near-infrared colours, like 2M1207b \citep[$J-K_s = 3.1$,][]{2004A&A...425L..29C}, PSO~J318.5338$-$22.8603 \citep[$J - K_s = 2.84$,][]{2013ApJ...777L..20L} and the HR8799 planets \citep{2008Sci...322.1348M}, and their photometric and spectroscopic properties are believed to be heavily influenced by the atmospheric condensates \citep[e.g.][]{2011ApJ...737...34M}. URLs can therefore be considered a ``bridge'' between brown dwarfs and giant planets atmospheres, thus a useful test-bed for the atmospheric models. Also, these objects can be used as probes to understand the physics of dust clouds, and to disentangle the effects of surface gravity and metallicity in the L type temperature regime.

In this contribution we present a new example of this class of objects, discovered in the United Kingdom Infrared Deep Sky Survey (UKIDSS) Large Area Survey (LAS): ULAS~J222711$-$004547. With a spectral type of L7pec (see Section 3), this object marks an extreme in the L/T transition, where the role of dust and condensates clouds becomes fundamental \citep[e.g.][]{2011ApJ...736...47B,2006ApJ...640.1063B,2001ApJ...556..357A}. The processes leading to the formation and subsequent disruption of said clouds, which determine the transition from L to T class objects, are still poorly understood. Constraining the effect of surface gravity and metallicity in these processes is an open challenge, with the different assumptions adopted in modern atmospheric models leading to significantly different results (especially for peculiar objects, see Section 6). ULAS~J222711$-$004547 represents a new opportunity to investigate this interesting stage in the evolution of sub-stellar objects.

In Section 2 we describe the candidate selection process. In Section 3 we present the spectrum obtained for our target, and discuss its peculiarities. In Section 4 we compare the photometry of ULAS~J222711$-$004547 with the population of known brown dwarfs. In Section 5 we calculate the target's proper motion, using the data available in the literature. In Section 6 we de-redden the spectrum of ULAS~J222711$-$004547 using extinction laws for different dust species, and we derive a typical grain size for the dust particles. In Section 7 we compare its spectrum with atmospheric models. Finally in Section 8 we summarize and discuss the results obtained.

\section{Candidate selection}
ULAS~J222711$-$004547 was selected from DR7 of the UKIDSS LAS, as part of a large spectroscopic campaign to constrain the sub-stellar birth rate \citep{2013MNRAS.430.1171D}. The sample included objects with declination $\leq$ 20 degrees, $J < 18.1$ and $Y - J > 0.8$. The initial list of candidate brown dwarfs was cross-matched against SDSS DR7 to identify objects with optical counter parts. The sample was then filtered to include only objects with $z - J \geq 2.4$ if $J- K \geq 1.0$ or $z - J \geq 2.9$ if $J- K < 1.0$. However, since mid-T dwarfs typically have $z - J > 3.0$ some objects would be too faint to be detected in SDSS. We therefore included these SDSS non-detections. Additional colour cuts were used and the list of quality flags applied to the sample can be found in \citet{2013MNRAS.430.1171D}, and we refer the reader to that contribution for a more detailed description of the candidate selection process. The final list of candidates was visually inspected to remove any possible mis-matches or spurious sources.

\section{Spectroscopic follow-up}
\subsection{Observations \& data reduction}
We obtained a medium resolution spectrum of ULAS~J222711$-$004547 using XSHOOTER on UT2 at the VLT. The instrument was used in echelle slit mode, covering the wavelength range of 0.3-2.5 $\mu$m. The full wavelength range is split into three separate arms, the UVB (0.3-0.55 $\mu$m), VIS (0.55-1.0 $\mu$m), and NIR (1.0-2.5 $\mu$m). The target was observed on the night of 2011-10-03 under the ESO program 088.C-0048 (P.I. A. C. Day-Jones), using a slit width of 1.0'' in the UVB arm and 0.9'' in the VIS and NIR arms. These correspond to a resolution of 5100 in the UVB arm, 8800 in the VIS arm and 5100 in the NIR arm. The object was visible in the 20 s acquisition image, and was placed on the centre of the slit. We took eight exposures in an ABBA nodded sequence, with individual exposure times of 230 s, 300 s and 390 s in the three arms. To allow for telluric correction, we observed the B7V star HIP014143 immediately after the target, matching the airmass of the observation.

The data were reduced using the ESO XSHOOTER pipeline (version 2.2.0). The pipeline removes non-linear pixels, subtracts the bias (in the UVB and VIS arm) or dark frames (in the NIR arm) and divides the raw frames by a master flat field. Biases, darks and flat fields are taken as part of the day time calibration. The images are then pair-wise subtracted to remove sky background. The pipeline then extracts the different orders in each arm, and rectifies them using a multi-pinhole arc lamp taken during the day. The individual orders are then corrected for the flexure of the instrument using a single-pinhole arc lamp, taken at night just before the target. Finally the orders are merged and the spectrum is flux calibrated using the spectrophotometric standard taken at the beginning of the night. The telluric standard was reduced in the same way, except that the sky subtraction is performed by fitting the background, as the telluric standard was not observed in nodding mode. 

We corrected for telluric absorption using our own IDL routine. First we removed the H lines from the telluric star by interpolating over them, and normalized it to one at 1.28 $\mu$m. Then we divided the standard by a black body curve for the appropriate temperature, normalized in the same way. Finally, we apply the telluric correction curve obtained this way to the target spectrum. 

The three arms were merged by matching the flux level in the overlapping regions between them. We checked the accuracy of the flux calibration by determining the synthetic UKIDSS MKO $YJHK$ magnitudes and comparing them with the measured ones from the ULAS. The agreement between synthetic and measured magnitudes is good, within the errors quoted in the ULAS. The spectrum was then binned in the wavelength direction to increase the signal-to-noise ratio (SNR), reducing the resolution by a factor of $\sim$10.

\begin{figure}
\includegraphics[width=8.5cm]{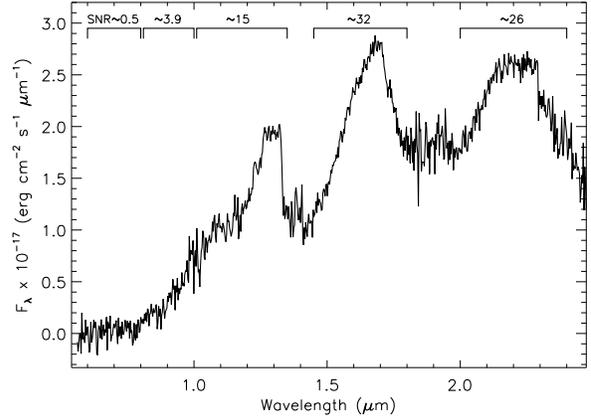}
\caption{The spectrum of ULAS~J222711$-$004547. At the top of the Figure we indicate the median SNR in each wavelength range. \label{spectrum}}
\end{figure}

The spectrum obtained for ULAS~J222711$-$004547 is plotted in Figure~\ref{spectrum}. We only show the 0.55-2.5 $\mu$m range, as we do not detect any flux at shorter wavelengths. The median SNR in the four main wavelength ranges (red-optical, $z$-band, $J$-band, $H$-band, and $K$-band) is indicated above the spectrum. Unlike normal L and T dwarfs, the spectrum peaks in the $H$-band, with the $K$-band reaching almost the same flux level. The optical and $J$-band are very smooth, and the CO absorption at 2.3 $\mu$m appears weaker than in other L/T transition dwarfs. The flux level at 2.2 to 2.28 $\mu$m is higher than in normal late-L dwarfs, an indication of a reduced Collision Induced Absorption (CIA) of H$_2$ or a reduced absorption from CH$_4$, the two major sources of opacity at these wavelengths \citep[e.g.][]{1999AJ....117.1010T,2003ApJ...596..561M}. A more detailed description of the spectral features of ULAS~J222711$-$004547 is given in the following sections.

\subsection{Spectral typing}
\label{spectral_typing}
\begin{figure}
\includegraphics[width=8.5cm]{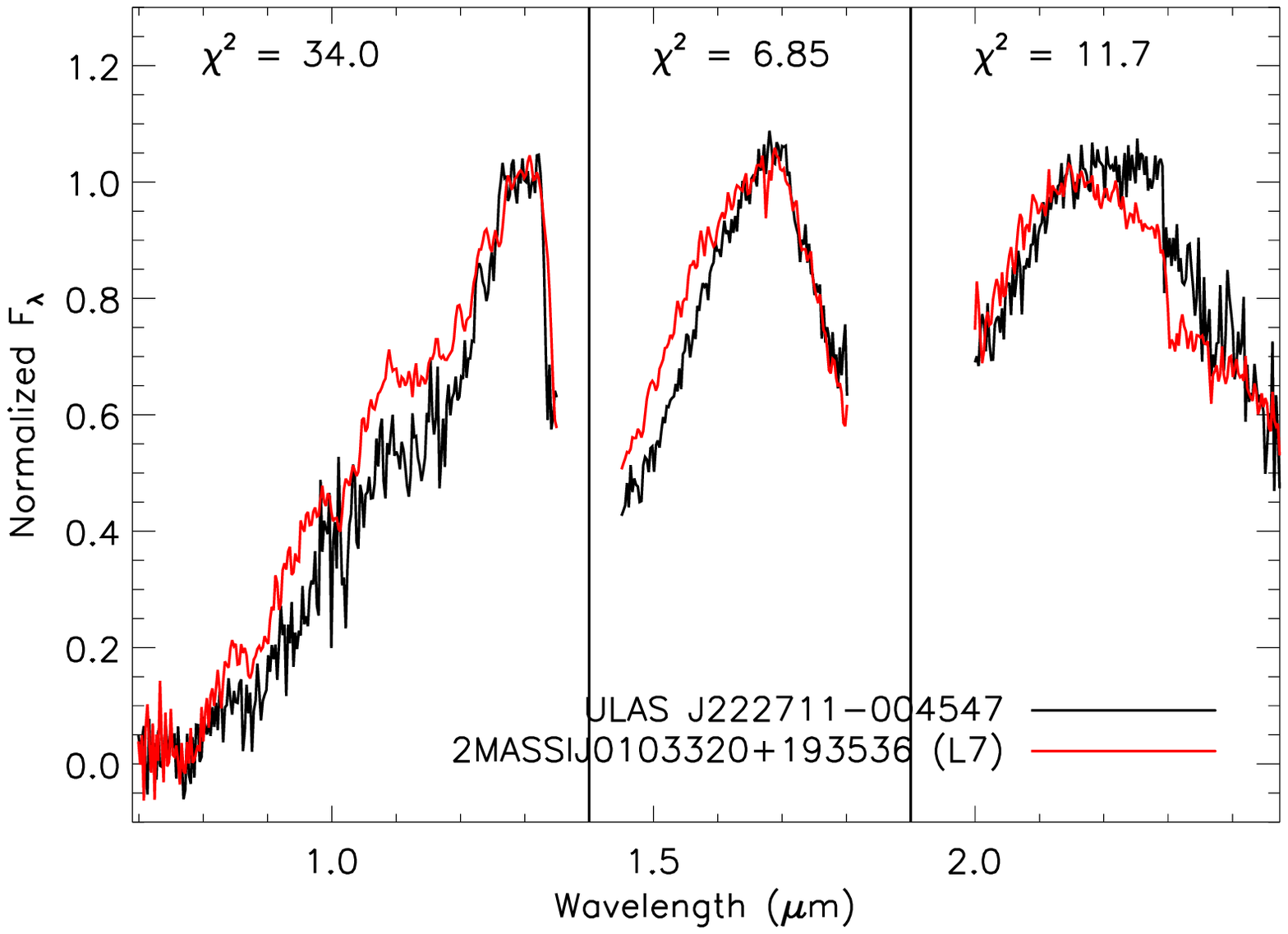}
\includegraphics[width=8.5cm]{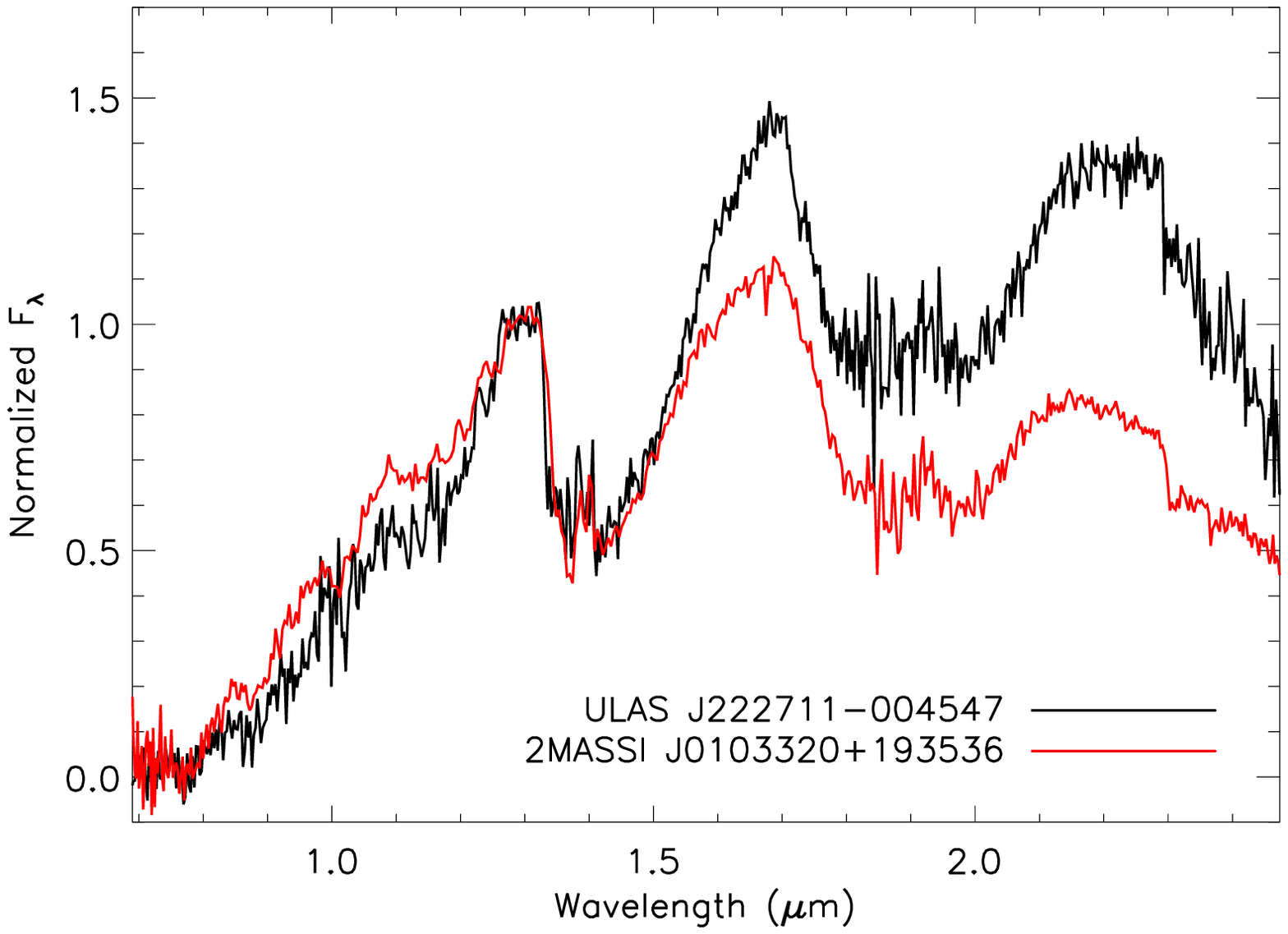}
\caption{The spectrum of ULAS~J222711$-$004547 (black) compared to the L7 standard 2MASSI~J0103320+193536 (red). \textit{Top:} both spectra are normalized to 1 in each portion. \textit{Bottom:} both spectra are normalized to 1 at 1.28 $\mu$m. \label{type}}
\end{figure}

Given the peculiarity of the spectrum of ULAS~J222711$-$004547, we cannot apply the standard spectral typing systems developed for L and T dwarfs \citep{1999ApJ...519..802K, 2006ApJ...637.1067B}. To assign a spectral type to our target we therefore split its spectrum into three separate portions: optical+$J$-band (0.7-1.35 $\mu$m), $H$-band (1.45-1.8 $\mu$m), and $K$-band (2.0-2.5 $\mu$m). We normalized the three portions separately to remove the steep red slope, and we compared them to the spectra of standard L and T dwarfs, treated the same way\footnote{All the spectra were taken from the Spex-Prism online library: http://pono.ucsd.edu/$\sim$adam/browndwarfs/spexprism}. In all three portions, the best fit is given by the L7 standard 2MASSI~J0103320+193536, but the match between the two spectra remains poor. We show the three portions of the spectrum with the best fit standard template on the top panel of Figure~\ref{type}. If we compare directly the spectrum of ULAS~J222711$-$004547 with 2MASSI~J0103320+193536 (normalizing the complete spectra to 1 at 1.28 $\mu$m) we note immediately that our target is much redder than the standard, with an increased flux level in the $H$ and $K$ bands, as we can see in the bottom panel of Figure~\ref{type}. We therefore classify ULAS~J222711$-$004547 as L7pec, following the notation described in \citet[Section 5.2]{2010ApJS..190..100K}.

\begin{figure}
\includegraphics[width=8cm]{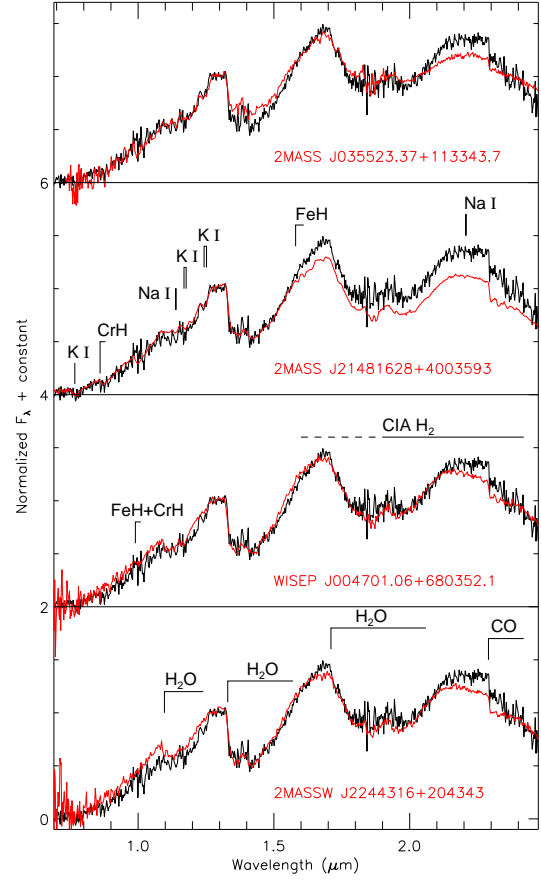}
\caption{The spectrum of ULAS~J222711$-$004547 compared to the known red L dwarfs 2MASS~J035523.37+113343.7 (L5$\gamma$), 2MASS~J21481628+4003593 (L6.5pec), WISEP~J004701.06$-$680352.1 (L7.5pec), and 2MASSW~J2244316+204343 (L7.5pec). All the spectra are normalized to 1 at 1.28 $\mu$m. \label{url}}
\end{figure}

A way to confirm its spectral type is to compare ULAS~J222711$-$004547 directly with known peculiar red L dwarfs. Figure~\ref{url} shows a comparison of its spectrum with the red L dwarfs 2MASS~J035523.37+113343.7, 2MASS~J21481628+4003593, WISEP~J004701.06$-$680352.1, and 2MASSW~J2244316+204343. The main absorption features are marked on the plot to ease the interpretation. Our target appears redder than all of the other peculiar L dwarfs. Its MKO $J-K$ is 2.79$\pm$0.06, thus is the reddest measured of any field brown dwarf (see Section 4). The L5$\gamma$ 2MASS~J035523.37+113343.7 looks very similar to our target in the optical and $J$-band, but has weaker water absorption bands at $\sim$1.4 and 1.9 $\mu$m, as expected in a earlier type dwarf. The $H$-band of ULAS~J222711$-$004547 looks less peaked than that of the L5$\gamma$, which could be an indication of higher surface gravity, but as pointed out by \citet{2013ApJ...772...79A} the ``peakiness'' of the $H$-band is not a wholly reliable indicator because it has been seen in dusty L dwarfs that are not young. Our target also appears much brighter than the L5$\gamma$ in the $K$-band, which is again expected in a later type dwarf. The L6.5pec 2MASS~J21481628+4003593 reproduces very well the shape and flux level of the optical and $J$-band of ULAS~J222711$-$004547. It also matches the relative depth of the water absorption bands, not surprisingly given the very small difference in spectral type. On the other hand, our target appears significantly redder in the $H$ and $K$-band, and with a weaker FeH absorption feature at $\sim$ 1.6 $\mu$m. These could be evidence for an enhanced dust content in the photosphere of ULAS~J222711$-$004547. The L7.5pec WISEP~J004701.06$-$680352.1 and 2MASSW~J2244316+204343 look similar to our target in the optical and $J$-band, but with less pronounced FeH absorption bands, a generally smoother continuum, and a slight flux excess between $\sim$0.9 and 1.1 $\mu$m. Both objects also show stronger H$_2$O absorption at $\sim$ 1.1 $\mu$m. All these differences are consistent with their slightly later spectral types. The $H$-band is well fitted, but ULAS~J222711$-$004547 presents a slight flux excess at the peak of the band. Our target is also brighter than both L7.5pec in the $K$-band, with a plateau in flux between $\sim$2.1 $\mu$m and the CO band head. An enhanced dust content of the photosphere and a slightly suppressed molecular hydrogen CIA in ULAS~J222711$-$004547 could be the cause of these discrepancies.

Using the XSHOOTER spectrum we also determined spectral indices for our target. The indices are those used for spectral classification, as defined in \citet{2002ApJ...564..466G} and \citet{2006ApJ...637.1067B}, and the gravity-sensitive indices defined in \citet{2013ApJ...772...79A}. The values obtained for ULAS~J222711$-$004547 are listed in Table~\ref{indices}, along with reference values for ``field'' and low gravity L7s (where available). Some of those indices are outside the defined ranges for spectral typing, except the H$_2$O 1.5 $\mu$m index which indicates a spectral type of L8, the CH$_4$ 2.2 $\mu$m which indicates a spectral type of L5, and the H$_2$O$-$H which indicates a spectral type of T0. We then applied the gravity classification method defined in \citet[ Table 9 and 10]{2013ApJ...772...79A}. The method is based on assigning ``scores'' of 0, 1 or 2 to different spectral features based on their strength. Specifically, the features considered are the FeH absorption bands at 0.998 and 1.2 $\mu$m (which strength is measured by the FeH$_z$ and FeH$_J$ indices respectively), the VO absorption band at 1.058 $\mu$m (measured by the VO$_z$ index), the alkali lines (measuring the equivalent width of the Na I line at 1.1396 $\mu$m and of the K I lines at 1.1692, 1.1778, and 1.2529 $\mu$m), and the ``peakiness'' of the $H$-band (measured by the $H$-cont index). The criteria to assign the scores are defined only for given spectral type ranges, and for some features do not extend down to L7. In those cases we assign a qualitative score based on a direct comparison of our target indices with the values plotted in \citet{2013ApJ...772...79A} for known low-gravity objects. We assign a score of 1 to the FeH absorption features based on the FeH$_J$ index only, as the FeH$_z$ feature falls in the very noisy region of the spectrum, at the edge of the VIS and NIR arms, where the flux calibration of our spectrum is less reliable. We assign a score of 0 to the VO feature as its index falls exactly on the line defining the average value for normal field dwarfs \citep[see][Figure 20]{2013ApJ...772...79A}. Finally, the alkali lines give an average score of 0 while the $H$-cont index gives a score of 2 \citep[based on the criteria defined in Table 9 and 10 of][]{2013ApJ...772...79A}. The median score for ULAS~J222711$-$004547 is therefore 0.5, and so we classify it as a field-gravity dwarf (hereafter FLD-G), i.e. the surface gravity is consistent with normal field dwarfs. This finding is in agreement with the results of \citet{2013ApJ...772...79A}, where they classified all the URLs in their sample as FLD-G dwarfs.

\begin{table}
\centering
\begin{tabular}{l c c c}
\hline
Index & \multicolumn{3}{c}{Value} \\
 & Target & Field L7 & Low gravity L7 \\
\hline
H$_2$O$-$J & 0.58$\pm$0.14 & $\gtrsim$ 0.70 & $\ldots$ \\
H$_2$O$-$H & 0.66$\pm$0.08 & $\gtrsim$ 0.70 & $\ldots$ \\
H$_2$O 1.5 $\mu$m & 1.71$\pm$0.23 & 1.65-1.70 & $\ldots$ \\
H$_2$O$-$K & 0.95$\pm$0.12 & $\gtrsim$ 0.70 & $\ldots$ \\
CH$_4$ 2.2 $\mu$m & 0.98$\pm$0.13 & 1.075-1.125 & $\ldots$ \\
FeH$_J$ & 1.00$\pm$0.17 & $\sim$1.1 & $\ldots$ \\
VO$_z$ & 0.97$\pm$0.15 & $\sim$0.97-1.00 & $>$ 1.00 \\
K I$_J$ & 1.02$\pm$0.10 & 1.01-1.07 & $<$ 1.01 \\
$H$-cont & 0.97$\pm$0.07 & $<$ 0.888 & $\geq$ 0.888 \\
\hline
Spectral & \multicolumn{3}{c}{Equivalent Width (\AA)} \\
Feature  & Target & Field L7 & Low gravity L7 \\
\hline
Na I 1.138 $\mu$m & 8.12$\pm$0.44 & $>$ 3.175 & $\leq$ 3.175 \\
K I 1.169 $\mu$m & 12.65$\pm$0.87 & $>$ 6.496 & $\leq$ 6.496 \\
K I 1.177 $\mu$m & 11.21$\pm$0.88 & $>$ 8.154 & $\leq$ 8.154 \\
K I 1.253 $\mu$m & 7.78$\pm$0.69 &  $>$ 4.545 & $\leq$ 4.545 \\
\hline
\end{tabular}
\caption{Spectral indices and equivalent widths for ULAS~J222711$-$004547. The indices are defined in \citet{2002ApJ...564..466G}, \citet{2006ApJ...637.1067B}, and \citet{2013ApJ...772...79A}. Where available, comparison values for ``field'' and low gravity L7s are given. \label{indices}}
\end{table}

\section{Photometric properties}
 We obtained the photometry of ULAS~J222711$-$004547 from the UKIDSS LAS, the 2MASS Point Source Catalogue, the \textit{WISE} All-Sky Data Release, and the \textit{NEOWISE} Post-Cryo Data Release. The object is undetected in 2MASS $J$-band, so we determined its synthetic magnitude in that band from the measured spectrum. We checked the accuracy of our synthetic value by comparing it with the magnitude obtained converting its UKIDSS MKO $J$-band magnitude using the equations presented in \citet{2004PASP..116....9S}. The two values are in good agreement with each other (18.27$\pm$0.06 vs. 18.25$\pm$0.04) and we therefore assume as our final 2MASS $J$-band magnitude 18.26$\pm$0.05. 

We also obtained a NTT/SOFI $J$-band image of ULAS~J222711$-$004547. The observations were made on the night of 2013-06-22 as part of the NPARSEC program (ESO 186.C-0756 P.I. R. L. Smart) and the observing strategy is fully described in \citet{2013MNRAS.433.2054S}. Briefly we observed with a nine point dither pattern of 4 $\times$ 30 s exposures (i.e. NDIT=4, DIT=30 s) for a total integration time of 18 minutes.  This was coadded using the {\it jitter} software, and centroids and instrumental magnitudes were found from the resulting coadded image using the {\it SExtractor} barycenter and psf fitting procedures \citep{1996A&AS..117..393B}. We calculated the zero point of the field using the measured 2MASS $J$-band magnitudes from the 2MASS Point Source Catalogue (hereafter PSC) of the reference stars in the field, converting them in ESO/SOFI magnitudes using the equations presented in \citet{2001AJ....121.2851C}. The SOFI $J$-band magnitude for ULAS~J222711$-$004547 is 18.13 $\pm$ 0.05. This value agrees with the 2MASS synthetic value, although only at the 2$\sigma$ level. This slight discrepancy could be due to the NTT/SOFI $J$-band filter bandpass including the telluric absorption bands, whose variability affects differently the early type stars used for photometric calibration and the brown dwarf.

There is no clear evidence for variability in the \textit{WISE} single exposures in \textit{W}1 and \textit{W}2, nor in the \textit{NEOWISE} single exposures in \textit{W}1 and \textit{W}2.

\begin{figure*}
\includegraphics[width=17cm]{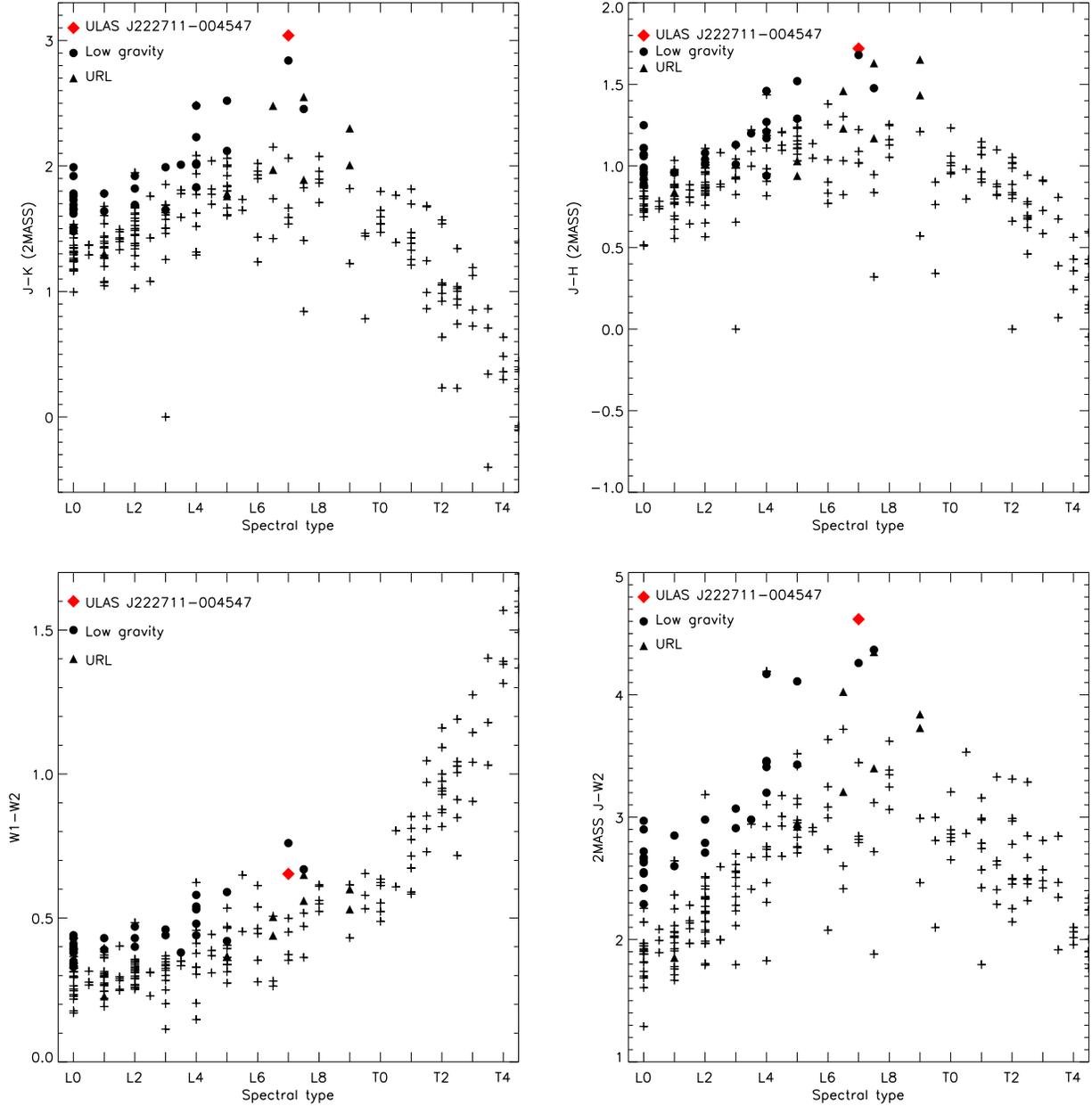}
\caption{Colour $-$ spectral type diagrams comparing the photometry of ULAS~J222711$-$004547 with other known L and T dwarfs. ``Normal'' field objects are plotted as crosses, known low-gravity dwarfs as circles, and unusually red L dwarfs (URLs) as triangles. ULAS~J222711$-$004547 is plotted as a red diamond. \label{photometry_1}}
\end{figure*}

\begin{figure*}
\includegraphics[width=17cm]{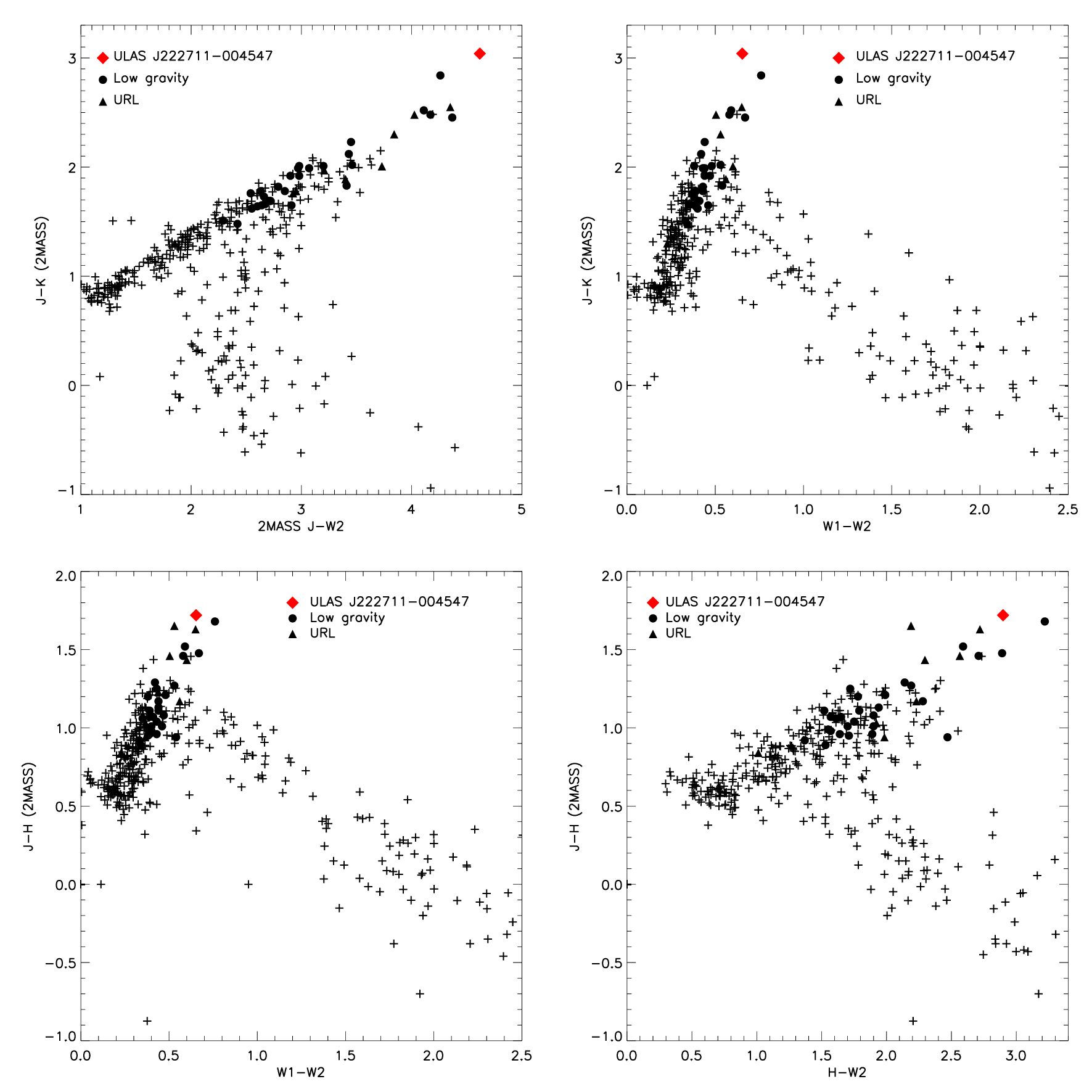}
\caption{Colour $-$ colour diagrams comparing the photometry of ULAS~J222711$-$004547 with other known L and T dwarfs. L dwarfs form a sequence running from bottom-left to top-right in each panel. Plotting symbols follow the same convention of Figure~\ref{photometry_1}. \label{photometry_2}}
\end{figure*}

In Figure~\ref{photometry_1} and \ref{photometry_2} we compare the photometry of ULAS~J222711$-$004547 with known ``normal'', low-gravity, and unusually red L dwarfs from the literature. We note that our target appears much redder than any of the known late-L dwarfs, and it is still redder than most of the other known low-gravity and unusually red L dwarfs (see Figure~\ref{photometry_1}). It is particularly interesting to notice in Figure~\ref{photometry_2}, that our target marks the end of the L-dwarf sequence (running from bottom-left to top-right in the four panels) further stressing its extreme nature, and that its photometry is very similar to the recently discovered free-floating planetary-mass L7 PSO~J318.5338$-$22.8603 \citep{2013ApJ...777L..20L}.

\section{Astrometry}
\label{astrometry}
\begin{figure*}
\includegraphics[width=17cm]{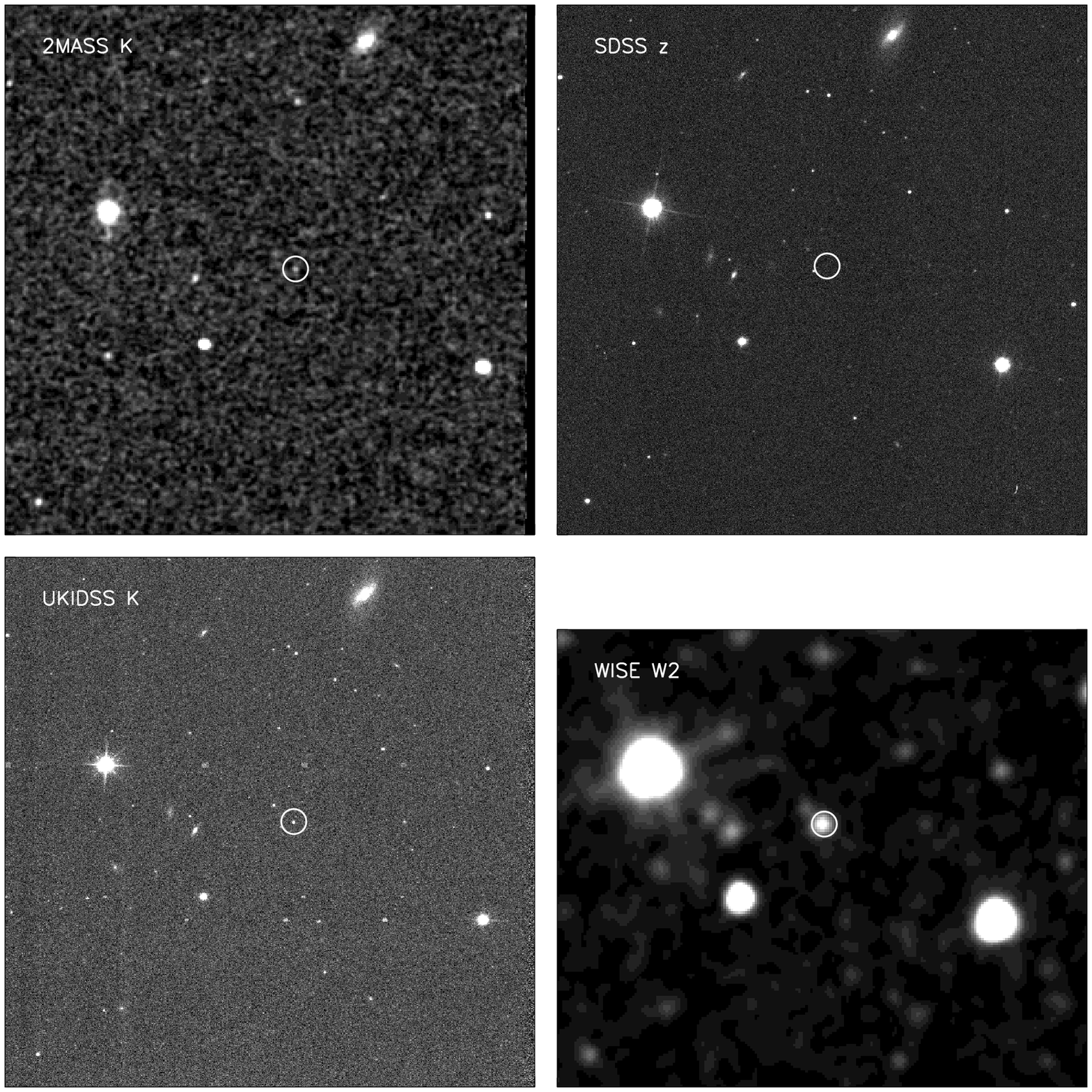}
\caption{Finder charts for ULAS~J222711$-$004547. The object position in each frame is marked with a circle. All images are oriented with north up and east left and are 5$\times$5 arcmin, except the \textit{WISE W}2 image which is 5$\times$4.3 arcmin. The target is undetected in the SDSS $z$-band. \label{brlt309_im}}
\end{figure*}

Other than in UKIDSS $YJHK$ bands, ULAS~J222711$-$004547 is detected in 2MASS $H$ and $K$-band, and in \textit{WISE} $W$1-$W$2-$W$3. In Figure~\ref{brlt309_im} we present the 2MASS $K$-band, UKIDSS $K$-band, and \textit{WISE} $W$2-band images of our target. We also show the SDSS $z$-band image of the field, but we note that ULAS~J222711$-$004547 is undetected. With a total baseline of more than 11 yr between the 2MASS and \textit{WISE} images, we can estimate the proper motion of our target. A linear fit to the measured positions in 2MASS, UKIDSS and \textit{WISE} gives a proper motion of $\mu_\alpha$\,cos\,$\delta$ = 100 $\pm$ 16 mas yr$^{-1}$ and $\mu_\delta$ = -30 $\pm$ 16 mas yr$^{-1}$. 

We tried to estimate the photometric distance to our target using the calibration published in \citet{2012ApJS..201...19D}. However, the peculiarity of ULAS~J222711$-$004547 makes the estimation unreliable, as the calibrations presented in the literature are applicable only to standard objects. It has indeed been pointed out by \citet{2012ApJ...752...56F} that unusually red L dwarfs tend to be underluminous compared to the average absolute magnitude of their spectral type by up to 1.0 mag. Moreover, their non-standard colours cause them to ``move'' up and down in an absolute magnitude $-$ spectral type plot. This results in the photometric distances given by the different magnitudes to be inconsistent with each other, spanning from 68 $\pm$ 12 pc if using MKO $M_{\rm Y}$ to 31 $\pm$ 5 pc when using \textit{WISE} $M_{W2}$. The value obtained using \textit{WISE} $M_{W2}$ should be the most accurate, as the flux at 4.6 $\mu$m should be much less impacted by dust.

We tested the possibility of the object being a member of one of the known young moving groups (hereafter MG) using the convergent-point method \citep[e.g.][and references therein]{2010MNRAS.402..575C,1999MNRAS.306..381D}. We de-composed the proper motion of ULAS~J222711$-$004547 into a component towards the convergent point of the MG ($\mu_\parallel$) and one perpendicular to that direction ($\mu_\perp$). For a MG member $\mu_\parallel$ must be positive while $\mu_\perp$ should ideally be zero. However, given the measurement errors on the proper motion of the target, and the intrinsic velocity dispersion of the MG, one should expect non-zero $\mu_\perp$. We calculated $\mu_\parallel$ and $\mu_\perp$ for our target relative to 13 known MGs, whose convergent points were taken from \citet{2013arXiv1303.5345M}. The results obtained are listed in Table~\ref{mg}. The large uncertainties on the proper motion components and the weak constraint on the distance to ULAS~J222711$-$004547 prevent us from drawing any firm conclusion. It appears however unlikely for the target to be member of any of the MGs considered, given its significant $\mu_\perp$. Even assuming the smaller distance estimate (31 $\pm$ 5 pc) the $\mu_\perp$ obtained corresponds in fact to peculiar velocities larger than 3 km s$^{-1}$, and generally above the 3$\sigma$ velocity dispersion of the MGs. Our results are in agreement with the Bayesian Analysis for Nearby Young AssociatioNs \citep[BANYAN,][]{2013ApJ...762...88M} online tool, that returns a probability of 97$\%$ of the object being a field ``old'' dwarf.

\begin{table}
\centering
\caption{Convergent point calculations for ULAS~J222711$-$004547. \label{mg}}
\begin{tabular}{l c c}
\hline
Group & $\mu_\parallel$ & $\mu_\perp$ \\
  & (mas yr$^{-1}$) & (mas yr$^{-1}$) \\
\hline
32 Ori & 101 & 26 \\
AB Dor & 89 & 53 \\
Alessi 13 & 102 & 23 \\
Argus & 100 & -28 \\
$\beta$ Pic & 100 & 27 \\
Carina-Near & 102 & -21 \\
Columba & 101 & 25 \\
Coma Ber & 98 & 36 \\
$\eta$ Cha & 97 & 37 \\
Hyades & 96 & -42 \\
Tucana & 100 & 29 \\
TW Hya & 102 & 24 \\
UMa (Sirius) & -65 & 82 \\
\hline 
\multicolumn{3}{l}{The uncertainty on $\mu_\parallel$ and $\mu_\perp$ is $\pm$16 mas yr$^{-1}$.}
\end{tabular}
\end{table}

Finally, we searched for common proper motion companions to ULAS~J222711$-$004547. We assumed the shortest distance estimate (31 $\pm$ 5 pc) and we looked for objects within 25000 AU from our target, with proper motion components within 2$\sigma$ from those of ULAS~J222711$-$004547. We used the PPMXL, LSPM, rNLTT, HIPPARCOS, TYCHO2 and UCAC4 catalogues, and we did not find any common proper motion companion to our target.

A summary of the properties of ULAS~J222711$-$004547 is given in Table~\ref{prop}, where we present coordinates, the available photometry, spectral type, proper motion, and the photometric distance range. 

\begin{table}
\centering
\caption{Summary of the properties of ULAS~J222711$-$004547. \label{prop}}
\begin{tabular}{c c}
\hline
Parameter & Value \\
\hline
RA (J2000) & 22:27:10.8 \\
Dec (J2000) & -00:45:47.3 \\
SDSS $z$ & $>$ 20.8 \\
MKO Y & 19.50 $\pm$ 0.11 \\
MKO J & 18.11 $\pm$ 0.06 \\
MKO H & 16.61 $\pm$ 0.03 \\
MKO K & 15.32 $\pm$ 0.02 \\
2MASS J & \textit{18.26 $\pm$ 0.05}$^a$ \\
ESO/SOFI J & 18.13 $\pm$ 0.05 \\
2MASS H & 16.54 $\pm$ 0.26 \\
2MASS K & 15.22 $\pm$ 0.16 \\
\textit{WISE W}1 & 14.295 $\pm$ 0.031 \\
\textit{NEOWISE W}1 & 14.332 $\pm$ 0.020 \\
\textit{WISE W}2 & 13.642 $\pm$ 0.041 \\
\textit{NEOWISE W}2 & 13.599 $\pm$ 0.031 \\
\textit{WISE W}3 & 12.283 $\pm$ 0.409 \\
\textit{WISE W}4 & $>$ 8.592 \\
Spectral type & L7pec \\
$\mu_\alpha$\,cos\,$\delta$ (mas yr$^{-1}$) & 100 $\pm$ 16 \\
$\mu_\delta$ (mas yr$^{-1}$) & -30 $\pm$ 16 \\
$d_{\rm phot}$ (pc) & 31 $-$ 68 \\
\hline
\multicolumn{2}{l}{Note: (a) Synthetic value, derived} \\
\multicolumn{2}{l}{from the measured spectrum.} \\
\end{tabular}
\end{table}

\section{De-reddening}
\label{dereddening}
The extremely red colours and spectra of red L dwarfs could be caused by thicker clouds \citep{2008ApJ...678.1372C}. Such thicker condensate clouds are believed to be associated with low surface gravity or high metallicity \citep[e.g.][]{2007ApJ...655.1079L,2008ApJ...686..528L,2009ApJ...702..154S}. We note that the extinction cross sections of condensate clouds are very sensitive to the characteristic size of the particles so in this context ``thicker clouds'' may mean a higher optical depth due to slightly larger particles rather than a structural difference in a cloud layer.

Low surface gravity is a sign of youth, and in L dwarfs is marked by triangular-shaped $H$-band spectra \citep[e.g.][]{2001MNRAS.326..695L}. In the case of ULAS~J222711$-$004547, the $H$-band looks indeed slightly triangular compared to the standard L7 (see Figure~\ref{type}). The effect is however not as strong as seen in known young objects like 2MASS~J03552337+1133437 \citep{2009AJ....137.3345C} and 2MASS~J01225093$-$2439505 \citep{2013ApJ...774...55B} . Other signs of youth, like the Li I absorption doublet at 6708 \AA{} and the H$\alpha$ emission at 6563 \AA{} are not seen in the spectrum of our target. Their absence however does not rule out the possibility that ULAS~J222711$-$004547 is a young field brown dwarf, as pointed out by \citet{2009AJ....137.3345C} and \citet{2008ApJ...689.1295K}. Also, the kinematics of this object does not match any of the known young moving groups (see Section \ref{astrometry}), and it is therefore unlikely that ULAS~J222711$-$004547 is very young.

\subsection{Checking for interstellar reddening}

\begin{figure}
\includegraphics[width=8cm]{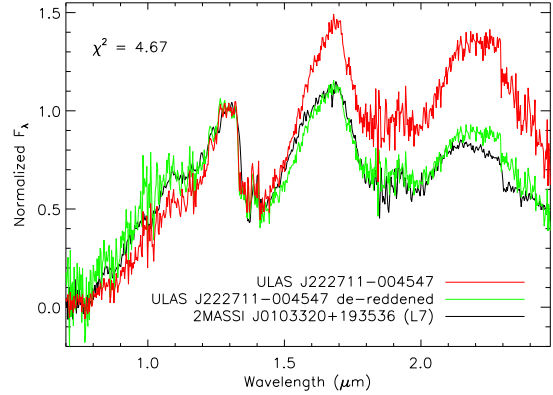}
\caption{The spectrum of ULAS~J222711$-$004547 (black) and its de-reddened version (green) compared to the L7 standard 2MASSI~J0103320+193536 (red). The spectrum has been de-reddened applying the \citet{1999PASP..111...63F} extinction curve with a colour excess of E(B-V)=1.1. \label{dered}}
\end{figure}

A way to determine the effect of dust clouds and their role in the formation of the spectra of peculiar red L dwarfs is to de-redden the spectra applying wavelength dependent corrections \citep[e.g.][]{2012AJ....144...94G}. We de-reddened the spectrum of ULAS~J222711$-$004547 using two different reddening curves: the \citet{1989ApJ...345..245C} and the more recent \citet{1999PASP..111...63F}, assuming the standard value $R(V) = 3.1$. For each reddening curve we selected the best colour excess E(B-V) $-$ standard template combination via $\chi^2$ minimization. In both cases the best fit is given by the L7 standard with a colour excess of E(B-V)=1.1. The results are shown in Figure~\ref{dered} where we compare the normalized spectrum of ULAS~J222711$-$004547 before and after the de-reddening with the standard L7 dwarf 2MASSI~J0103320+193536. Since there is no significant difference between the results obtained with the two different extinction laws, we only show the correction obtained with the \citet{1999PASP..111...63F} curve. The de-reddened version of the spectrum resembles closely to the L7 standard, but still shows a more peaked $H$-band, due to its blue-wing being fainter, i.e. the water absorption at 1.4 $\mu$m being stronger, than in 2MASSI~J0103320+193536. The $K$-band flux is also slightly enhanced, especially in its red end. We note in particular a ``bump'' between 2.1 and 2.3 $\mu$m, and a weaker CO absorption compared to the L7 standard. This result is quite remarkable and surprising when considering how different the interstellar dust and the atmospheric clouds of brown dwarfs are. 

There is of course the possibility that the reddening of the spectrum is due to interstellar dust clouds. However this is improbable as the object is at a distance of less than 70 pc (see Section 5) and that the asymptotic reddening for the field is E(B-V) = 0.068, which corresponds to an integrated extinction of approximately 0.061 mag, 0.039 mag, and 0.025 mag in the $J$, $H$, and $K$ band respectively. Reddening and extinction for the field were obtained using the Galactic Dust Reddening and Extinction service\footnote{http://irsa.ipac.caltech.edu/applications/DUST/}, which is based on the results of \citet{1998ApJ...500..525S}. We note that the value of E(B-V) measured for the field is almost constant, with $\Delta$ E(B-V) = 0.01 over an area of 2$\times$2 square degrees. This rules out peculiar properties of the interstellar medium and the presence of molecular clouds along the line of sight, and therefore justifies the use of $R(V) = 3.1$.

\subsection{De-reddening with dust typical of L dwarfs}

The surprisingly good results given by the method described above, motivated us to calculate more specific extinction curves, for the dust species that are typical of the atmospheres of brown dwarfs. In this contribution we have considered corundum (Al$_2$O$_3$), enstatite (MgSiO$_3$), and iron, as they are thought to be the most abundant in late-L dwarfs \citep[e.g.][and references therein]{2012ApJ...756..172M}. For each dust species we calculated the extinction cross section as a function of wavelength using an adapted version of the Mie scattering code of \citet{1983asls.book.....B}. Following \citet{2000ApJ...540..504S}, optical constants were taken from \citet{1995A&A...300..503D}, for enstatite; \citet{1995Icar..114..203K}, for corundum; and \citet{1985ApOpt..24.4493O}, for iron. For corundum we used the ``ISAS'' particles given in \citet{1995Icar..114..203K}, since this yields a single scattering albedo spectrum consistent with that shown by \citet{2000ApJ...540..504S}.

We have calculated the extinction for a range of characteristic grain radii (hereafter $r$) from 0.05 to 1.00 um. For each $r$ we used a Gaussian size distribution with a width $\sqrt{2} \sigma = 0.1 \times r$. This size distribution is wide enough to smooth over the interference effects that arise in Mie scattering. In reality, there is likely to be a broader distribution of grain sizes but the extinction curve is dominated by grains close to $r$, so it is useful to consider a characteristic size, even though it may not dominate the mass. For example, the scattering cross section of particles much smaller then the wavelength is proportional to the sixth power of the grain size (due to the Rayleigh-like fourth power in efficiency, combined with the physical cross section increasing with the square of radius). Therefore the size distribution would have to be very steep for the smaller particles to contribute significantly.

The range of $r$ that we consider is constrained by basic considerations of light scattering physics and by the fact that particles larger than 1 $\mu$m are expected to drop out of the photosphere. If the size parameter $x = 2 \pi r / \lambda$ of optically dominant grains is much bigger than 1 in the wavelength range under consideration, then the extinction is expected to be ``grey'', with little or no reddening effect. If $r < 0.05$ $\mu$m then it approaches the small grain limit, and no changes to the extinction curve would be expected if the grains were smaller.

\begin{figure}
\includegraphics[width=8cm]{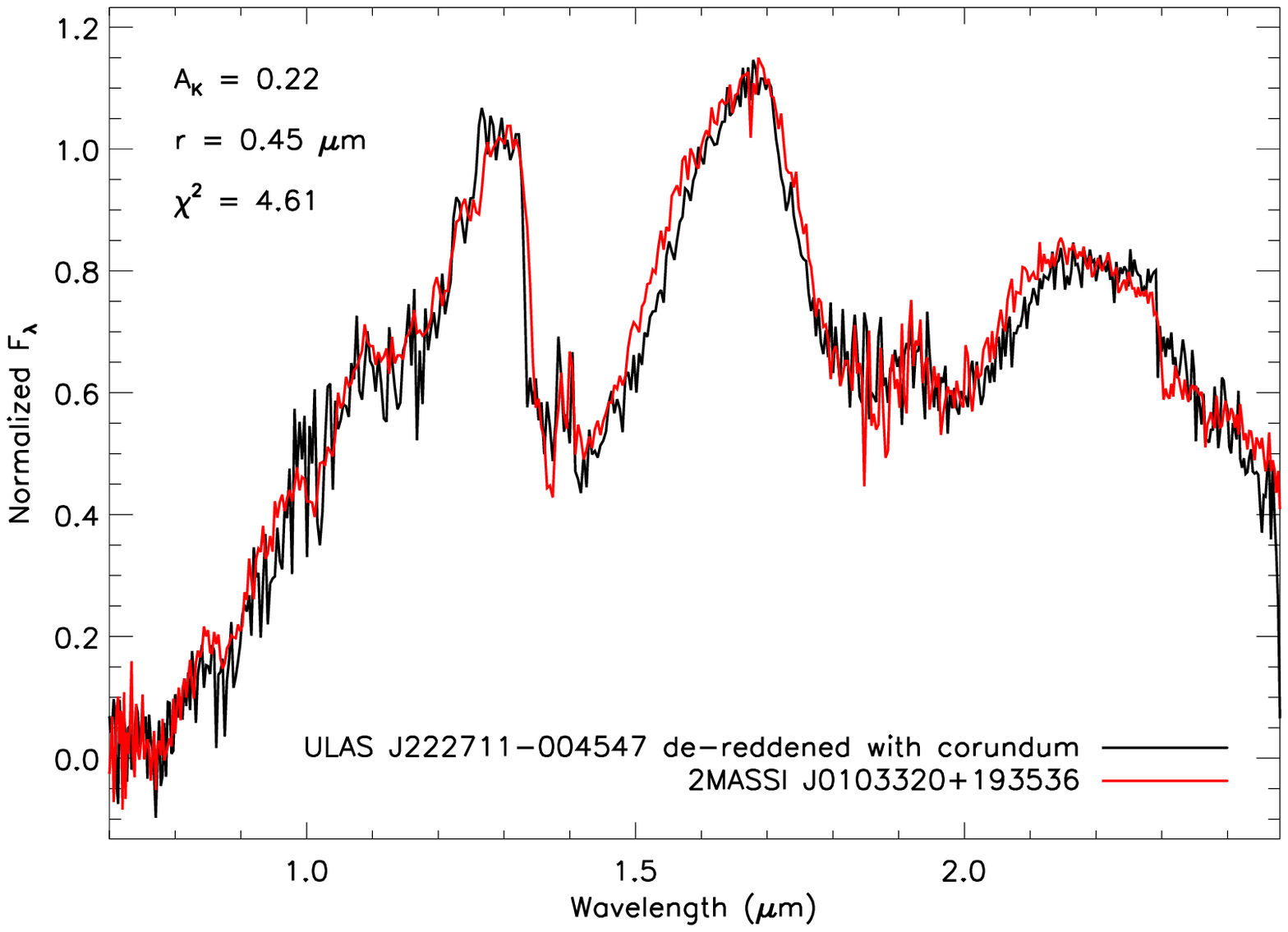}
\includegraphics[width=8cm]{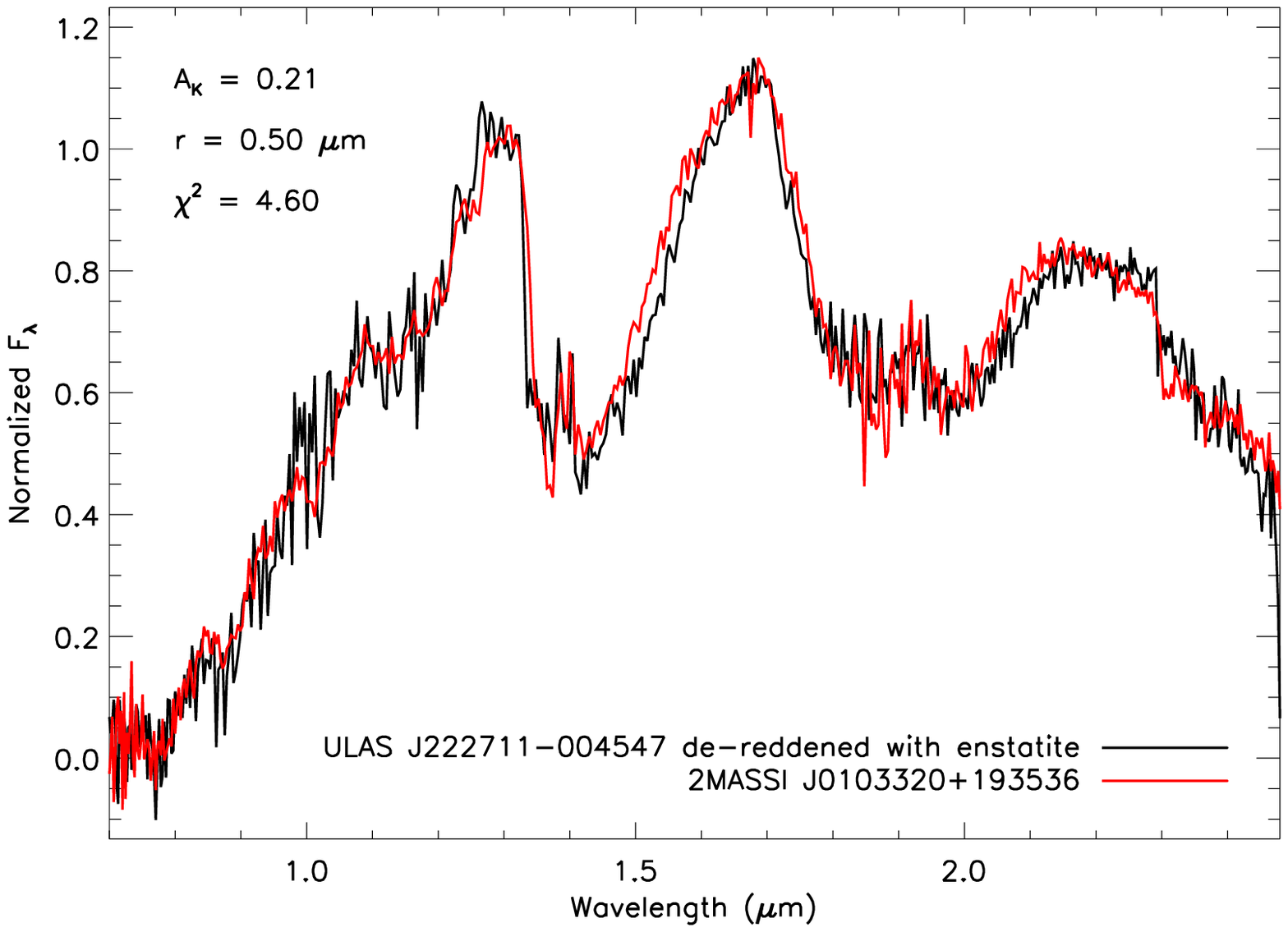}
\includegraphics[width=8cm]{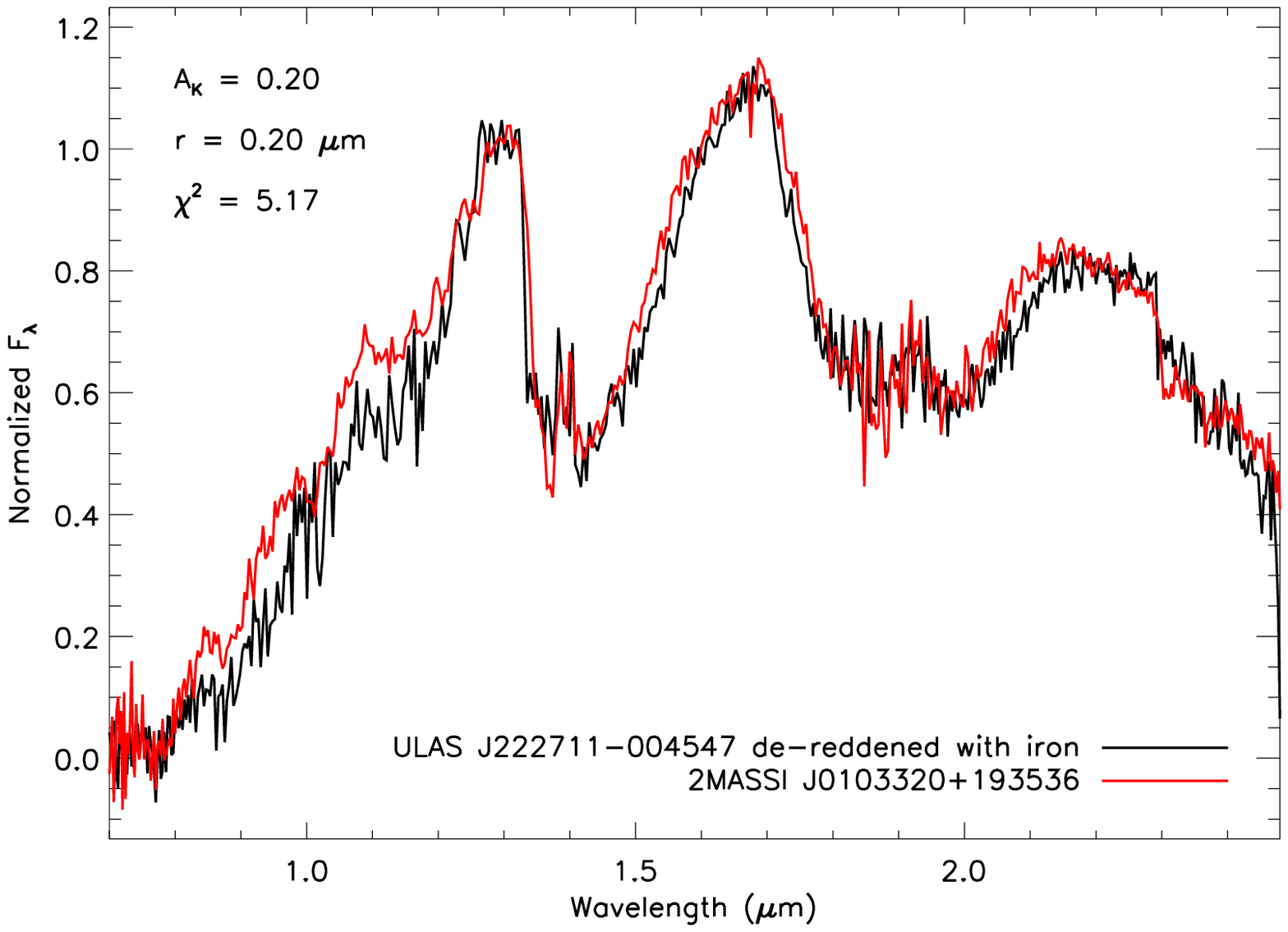}
\caption{The spectrum of ULAS~J222711$-$004547 (black line) de-reddened using extinction curves for corundum (top), enstatite (middle) and iron (bottom) compared to the L7 standard 2MASSI~J0103320+193536 (red line). The best fit parameters are shown in the top left corner of each panel. \label{dust}}
\end{figure}

We then de-reddened the spectrum of ULAS~J222711$-$004547 using the extinction curves and fit it to the already mentioned L7 spectroscopic standard 2MASSI~J0103320+193536. The fit has two free parameters: other than the grain size $r$, the normalization of the extinction curve at 2.20 $\mu$m (i.e. $A_{\rm K}$). In Figure \ref{dust} we show the best fit obtained for each dust species. Corundum and enstatite (top and middle panel of Figure \ref{dust}) give a very good fit for typical grain sizes of 0.45-0.50 $\mu$m. The de-reddened spectrum matches almost perfectly the standard, with only slight discrepancies in the CO absorption band at 2.3 $\mu$m and in the CrH band at 0.86 $\mu$m. The best fit grain size for iron is 0.20 $\mu$m (bottom panel of Figure \ref{dust}), but the quality of the fit is significantly poorer. While the NIR portion is quite well matched, the optical and J-band appear much fainter than in the standard. All the dust species give a best-fit $A_{\rm K}$ of 0.20-0.22, which corresponds to $\tau_{\rm K}$ = 0.18-0.20. The de-reddened \textit{WISE} photometry gives a \textit{W}1$-$\textit{W}2 = 0.62 (using the corundum or enstatite extinction curve), and 0.63 (using the iron extinction curve). The \textit{WISE} colours are still redder than the average even after de-reddening them, because the extinction curves we derived decline steeply as a function of wavelength, and the difference between the extinctions at 3.4 and 4.6 $\mu$m is therefore negligible. 

A complication is that there is a strong peak in the extinction cross section when the grain size is a little smaller than the wavelength of observation \citep[e.g. Figure 15 of][]{1998MNRAS.299..699L}. The strength of this peak and the corresponding size parameter depend on the shape and refractive index of the particles. This has the consequence that grains that are a bit smaller than the maximum size can have a significant effect at shorter wavelengths. For this reason we have attempted to de-redden with a power law size distribution of iron grains, in order to determine whether the poor fit to iron is due to the unusual optical properties of this metal or to the fact the large refractive index of iron causes smaller grains in the distribution to have a more noticeable effect. We determined extinction curves for exponents between 0 and -7.00 with steps of 0.25. We assume a fixed minimum grain size of 0.05 $\mu$m, while the maximum grain size $r_{\rm max}$ at which we truncate the power law is a parameter of the fit (as well as $A_{\rm K}$ and the exponent of the power law).

\begin{figure}
\includegraphics[width=8cm]{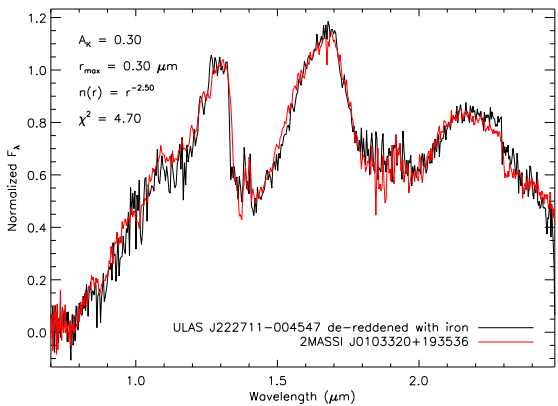}
\caption{The spectrum of ULAS~J222711$-$004547 (black line) de-reddened using the extinction curve for iron assuming a power law grain size distribution. The L7 standard 2MASSI~J0103320+193536 is plotted in red. The best fit parameters are shown in the top left corner. \label{iron_pl}}
\end{figure}

The best-fit obtained is shown in Figure \ref{iron_pl}. The maximum grain size obtained is 0.30 $\mu$m with $A_{\rm K}$ = 0.30 and a power law index of -2.50. The quality of the fit obtained is equal to that obtained with corundum and enstatite, and much better than the fit obtained assuming a narrow gaussian distribution around the characteristic $r$. This indicates that given its large refractive index, the size distribution of iron cannot be neglected when computing its extinction.

\subsection{Testing the method on other Unusually Red L dwarfs}

\begin{figure}
\includegraphics[width=8.5cm]{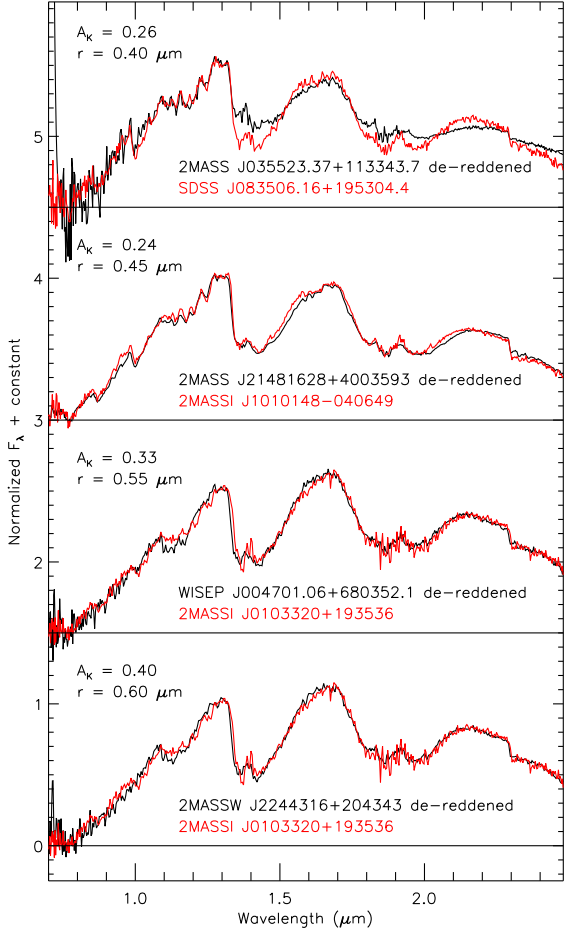}
\caption{The spectra of 2MASS~J035523.37+113343.7, 2MASS~J21481628+4003593, WISEP~J004701.06$-$680352.1, and 2MASSW~J2244316+204343 (from top to bottom), de-reddened with corundum. Overplotted in red are the corresponding standard templates used. The best fit parameters for each object are indicated on the left-hand side. \label{dered_urls}}
\end{figure}

To test the reliability of our method, we applied it to the other URLs plotted in Figure \ref{url}. An example of the fits obtained is shown in Figure \ref{dered_urls} where we plot 2MASS~J035523.37+113343.7, 2MASS~J21481628+4003593, WISEP~J004701.06$-$680352.1, and 2MASSW~J2244316+204343 (from top to bottom), dereddened with corundum. Overplotted in red are the corresponding standard templates used. WISEP~J004701.06$-$680352.1, and 2MASSW~J2244316+204343 give results that are very similar to those obtained for ULAS~J222711$-$004547. The best-fit grain size for corundum and enstatite is slightly larger (r = 0.50-0.60 $\mu$m for $A_{\rm K} \sim$ 0.5) and the quality of the fit to the L7 standard is surprisingly good. De-reddening using iron gives a best-fit grain size of 0.20 $\mu$m for both objects ($A_{\rm K} \sim$ 0.3), but the quality of the fit to the red-optical part of the spectrum is again poorer. The best-fit parameters for 2MASS~J21481628+4003593 against the L6 standard are identical to those given by our target, i.e. r = 0.45-0.50 $\mu$m with $A_{\rm K}$ = 0.22-0.24 for corundum and enstatite, and r = 0.20 $\mu$m and $A_{\rm K}$ = 0.20 using iron. The quality of the fit to the standard in this case is very good also for the de-reddening using iron, with a very good match to the entire spectrum. The consistency of these results strengthens the validity and the significance of our approach, further highlighting the importance of dust in the photosphere of URLs.

The fits obtained for 2MASS~J035523.37+113343.7 against the L5 standard are poorer. The best-fit grain sizes obtained for corundum, enstatite and iron are comparable to those obtained for the other URLs (0.40, 0.50, and 0.15 $\mu$m respectively, with $A_{\rm K}$ = 0.26, 0.32, and 0.19). However, the de-reddened spectrum resembles closely that of the standard in the red-optical and $J$-band, but the standard has stronger H$_2$O and CO absorption, a less peaked $H$-band, and it is brighter in the $K$-band. The poorer results obtained for this object could be due to a combination of two factors. First, 2MASS~J035523.37+113343.7 is classified as very-low gravity in \citet{2013ApJ...772...79A}, and it is likely to be a member of the AB Doradus moving group \citep{2013AJ....145....2F}. Therefore, to fully explain its spectral peculiarities low gravity cannot be neglected, and a simple dust de-reddening is not sufficient. Second, this object is of a slightly earlier spectral type compared to the other URLs considered, and the role of dust clouds in its photosphere is intrinsically less dominant.

\subsection{De-reddening using different templates}

\begin{figure}
\includegraphics[width=8.5cm]{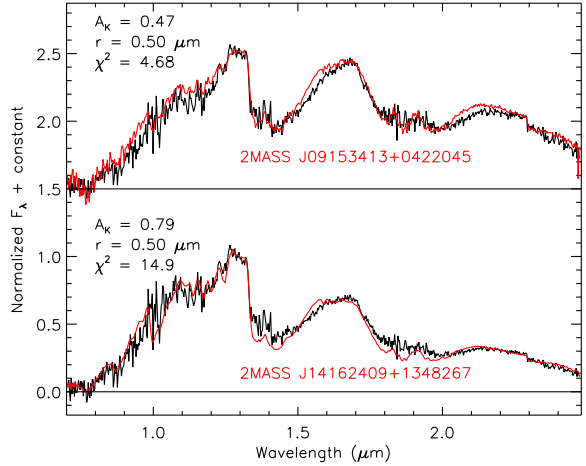}
\caption{The spectrum of ULAS~J222711$-$004547 de-reddened with corundum, compared to the slightly blue L7 2MASS~J09153413+0422045 (top), and the d/sdL7 2MASS~J14162409+1348267 (bottom). The best fit parameters for each fit are indicated on the left-hand side. \label{dered_blue}}
\end{figure}

We have also de-reddened the spectrum of ULAS~J222711$-$004547 using different comparison dwarfs. Instead of the L7 standard, we used the L7 2MASS~J09153413+0422045, which is slightly bluer than the standard, and the d/sdL7 2MASS~J14162409+1348267, which is a known slightly metal-poor dwarf \citep{2010ApJ...710...45B,2010AJ....139.1045S}. The fit to 2MASS~J09153413+0422045 gives typical grain sizes slightly larger but consistent with those obtained with the fit against the L7 standard. We obtain r = 0.50, 0.60, and 0.20 $\mu$m for corundum, enstatite and iron respectively. The main difference between the two fits is, not surprisingly, the larger $A_{\rm K}$ given by the bluer template, in the range 0.35 to 0.60. The quality of the fit is still extremely good for the entire spectrum. When fitting the d/sdL7 2MASS~J14162409+1348267 we obtain the same typical grain sizes, but a higher extinction $A_{\rm K}$ = 0.79, 0.95, and 0.56 for corundum, enstatite and iron respectively. The quality of the fit remains quite good for corundum and enstatite, except for the $H$-band peak that appears too triangular compared to the d/sdL7. De-reddeing with iron gives a very poor fit at wavelengths shorter than 1.2 $\mu$m, and a good fit to the rest of the spectrum. The results are shown in Figure \ref{dered_blue} where we show the spectrum of ULAS~J222711$-$004547 de-reddened using the corundum extinction curve. The two different standards that we used are overplotted in red for comparison.

These surprisingly good results suggest that the differences in photometry and spectra seen in objects of similar spectral types \citep[especially at late-L type, e.g.][]{2013EPJWC..4706007Z} could be explained almost entirely by assuming differences in the optical thickness and depth of the cloud deck.\\[\baselineskip]

The results of this de-reddening however have to be taken with caution. In fact we do not use the extinction curves in a defined structure or medium but simply apply them directly to the observed spectrum. This is equivalent to adding a layer of cold dust at the top of the brown dwarf's photosphere, neglecting the effect of scattering (i.e. both scattered and absorbed photons are considered to disappear), and without taking into account any effect that the extra dust would cause on the atmosphere of the object (e.g. elements depletion, backwarming effect, etc.). Physically, the dust layer as fitted here could be located anywhere above the photosphere, and the fitted extinction $A{\rm _K}$ represents the effect of this dust layer over and above the normal extinction effect of dust in the spectroscopic standard. More realistically, at least some of the additional dust in the red L dwarf is likely to be mixed in with the photosphere, which would mean that the $A{\rm _K}$ parameter is under-estimated. In fact, it is not necessary for the dust in red L dwarfs to be located at a different altitude than in normal L dwarfs. It is only necessary for it to have higher optical depth, perhaps due to larger grain size, or a higher space density of particles, rather than a physically thicker cloud layer.

A larger typical dust grain size in the photosphere can naturally arise from lower gravity, which would increase the maximum grain size that can remain suspended in the photosphere. However, in ULAS~J222711$-$004547 there is no clear evidence for low gravity (see Section \ref{spectral_typing}), so the explanation could be a higher than solar metallicity.

\section{Model fitting}
\begin{figure}
\includegraphics[width=8.5cm]{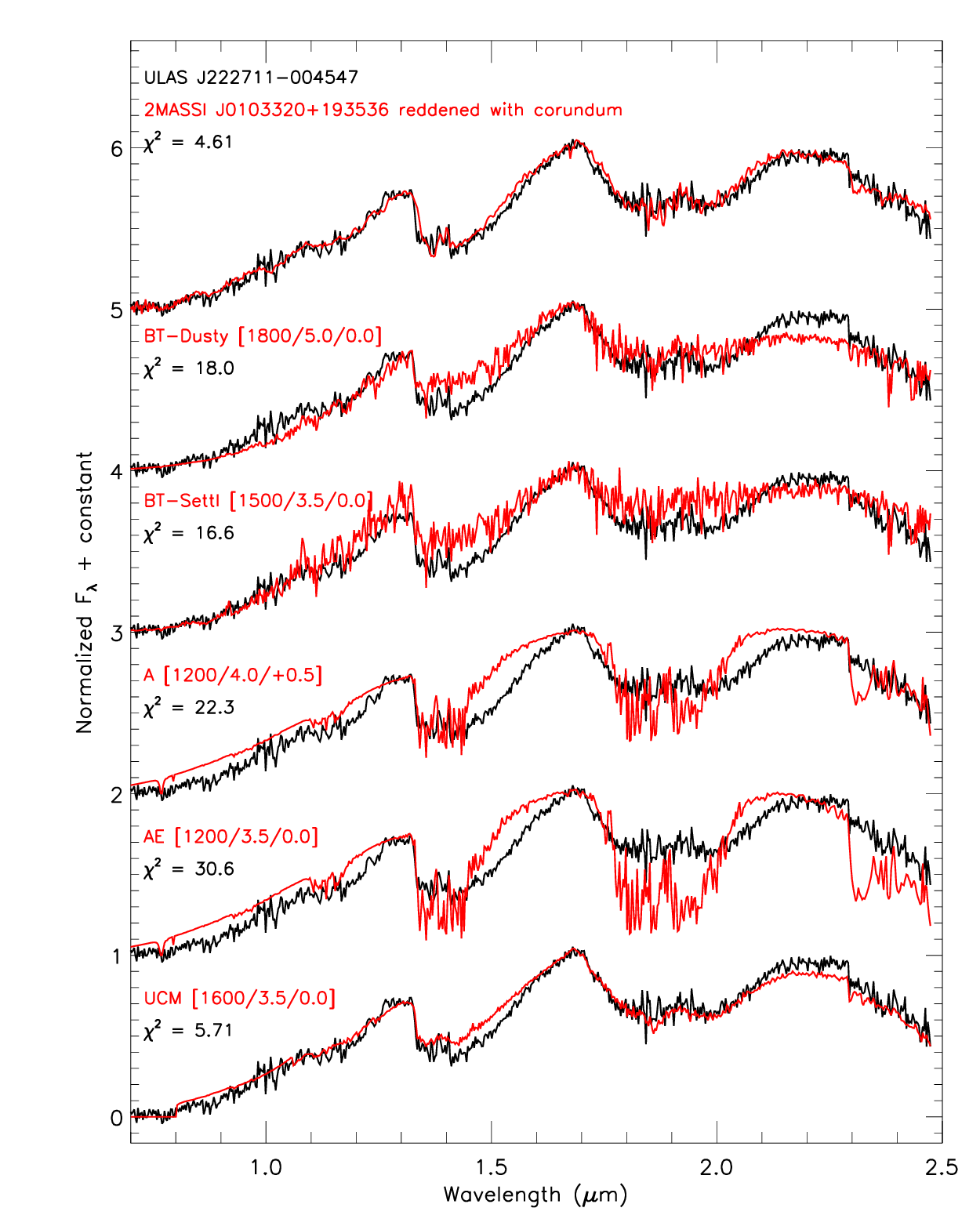}
\caption{The spectrum of ULAS~J222711$-$004547 compared to atmospheric models. The models used are the BT-Settl and BT-Dusty \citep{2011ASPC..448...91A}, the A and AE models \citep{2011ApJ...737...34M}, and the UCM models \citep{2002ApJ...575..264T,2004ApJ...607..511T,2005ApJ...621.1033T}. For each model we indicate in bracket effective temperature, surface gravity and metallicity, following the scheme [$T_{\rm eff}$/log(g)/[Fe/H]]. For comparison, we show at the top of the plot the fit given by the L7 standard reddened with our corundum extinction curve.\label{models}}
\end{figure}

\begin{figure}
\includegraphics[width=8.5cm]{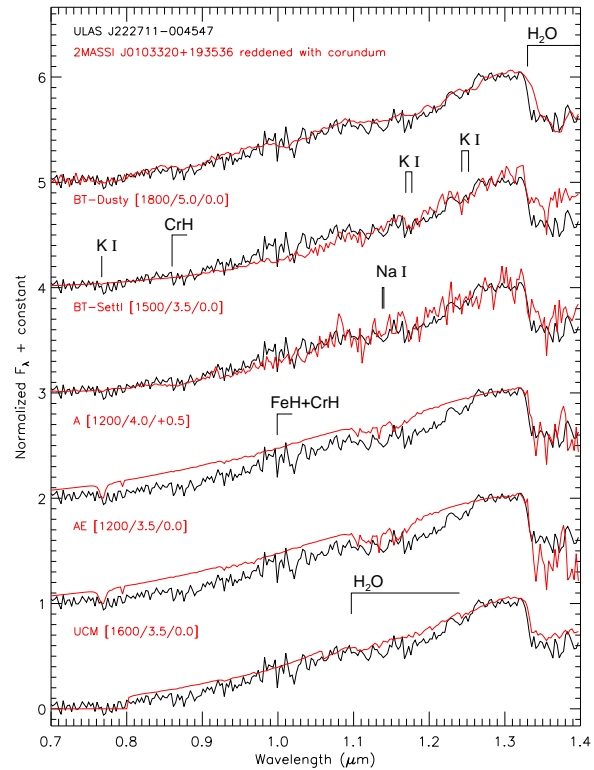}
\caption{A zoom into the optical+$J$-band portion of the spectrum. The models overplotted are the same as in Figure~\ref{models}. \label{models_J}}
\end{figure}

\begin{figure}
\includegraphics[width=8.5cm]{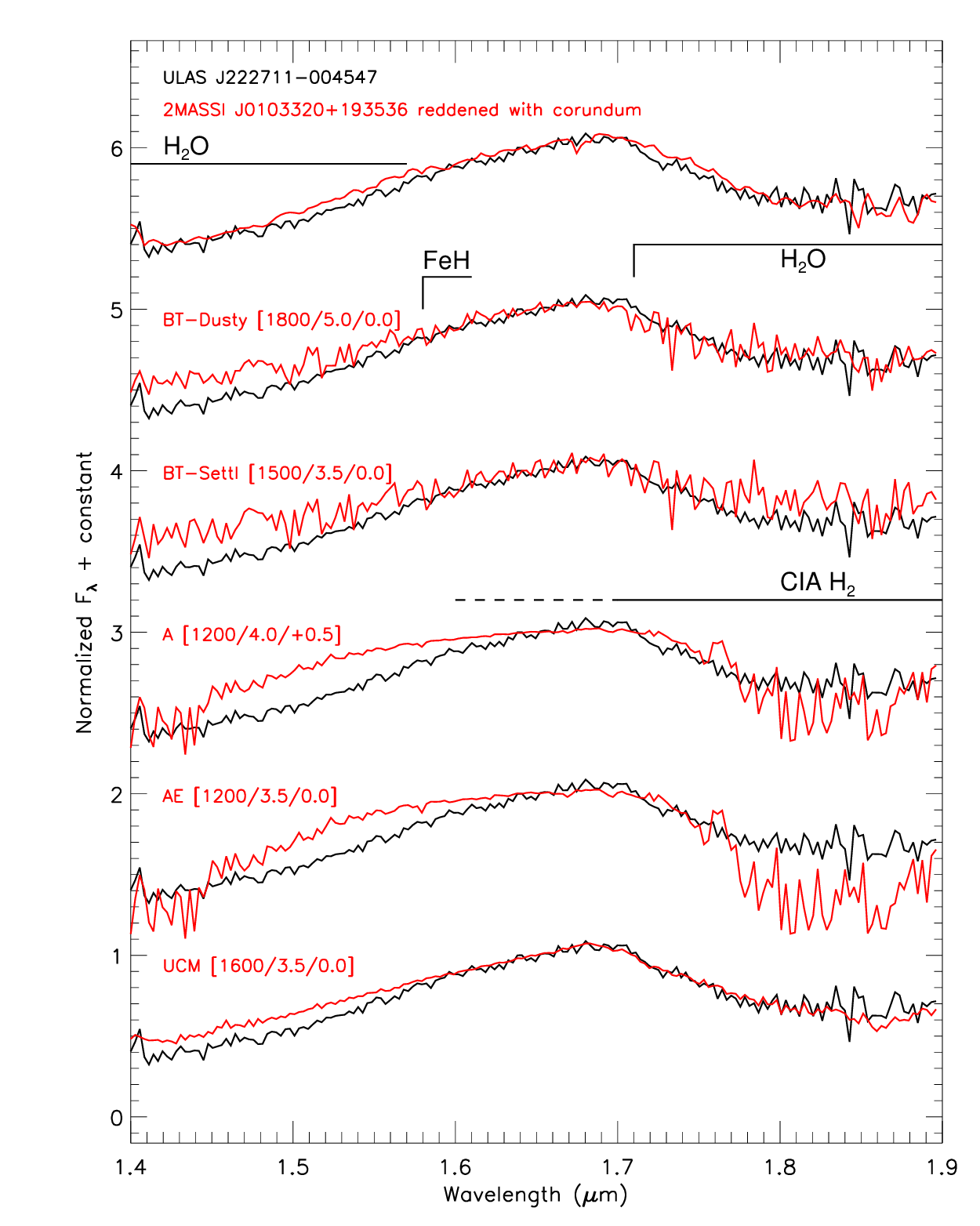}
\caption{A zoom into the $H$-band portion of the spectrum. The models overplotted are the same as in Figure~\ref{models}. \label{models_H}}
\end{figure}

\begin{figure}
\includegraphics[width=8.5cm]{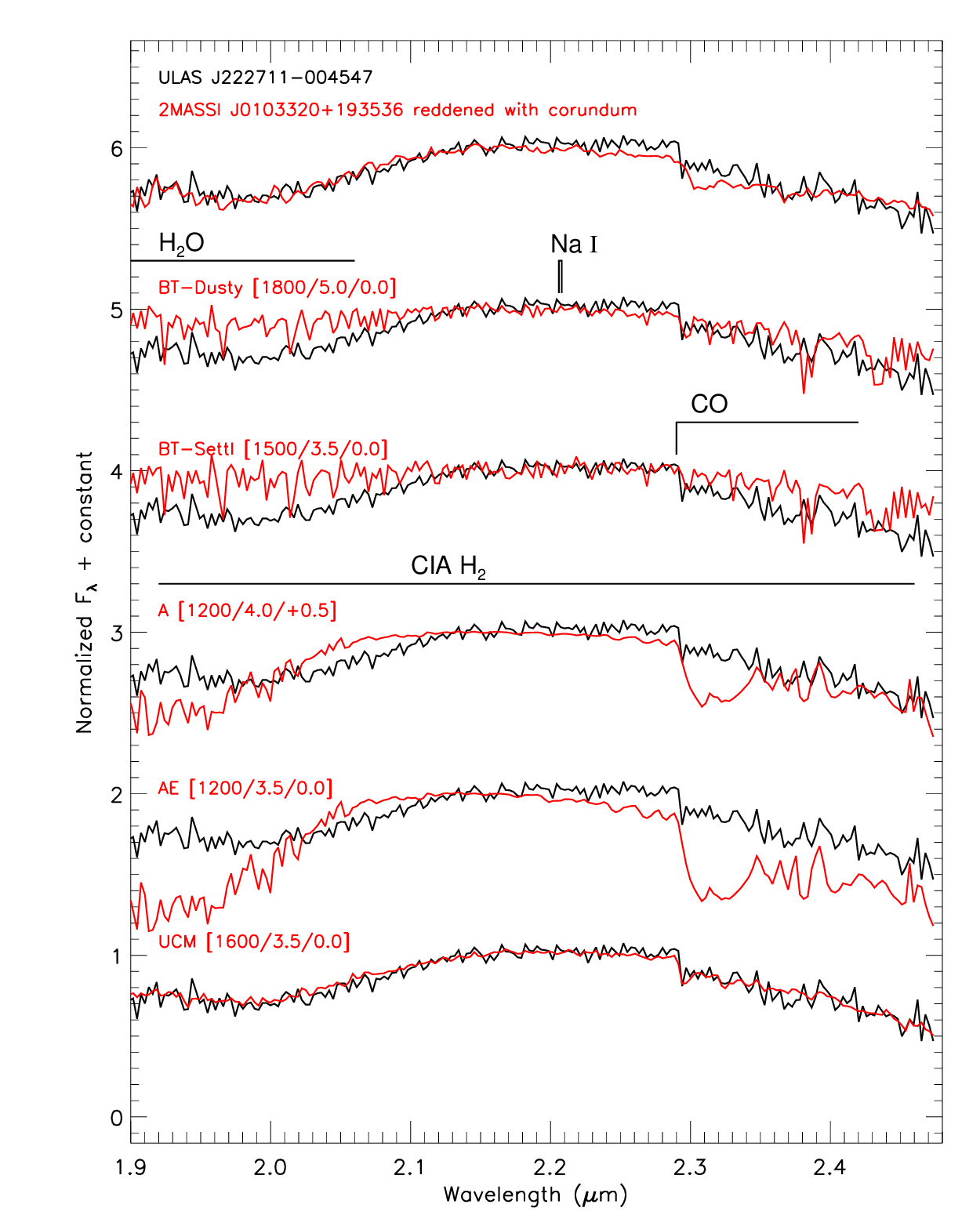}
\caption{A zoom into the $K$-band portion of the spectrum. The models overplotted are the same as Figure~\ref{models}. \label{models_K}}
\end{figure}

We fit the spectrum of ULAS~J222711$-$004547 with a set of atmospheric models to try to understand the origin of its very red infrared colours. We used the BT-Settl and BT-Dusty models from \citet{2011ASPC..448...91A}, the A and AE models from \citet{2011ApJ...737...34M}, and the Unified Cloudy Models (hereafter UCM) from \citet{2002ApJ...575..264T,2004ApJ...607..511T,2005ApJ...621.1033T}. For each set of models, we selected the best fit one via $\chi^2$ fitting. The results are shown in Figure~\ref{models} where we present the fit to the entire spectrum, and in Figure~\ref{models_J}, \ref{models_H} and \ref{models_K}, where we show a zoom to the optical+$J$-band, $H$-band, and $K$-band respectively (separately normalized to 1 at their peaks).  At the top of each Figure, we plot for comparison the fit given by the L7 standard reddened using the extinction curve for corundum derived in Section \ref{dereddening}.

\subsection{BT-Dusty and BT-Settl models}

In the BT-Dusty and BT-Settl models dust species abundances are calculated assuming chemical equilibrium. In the BT-Dusty models, no gravitational settling is considered, and the clouds are homogeneously distributed in the atmosphere of the brown dwarf. In the BT-Settl models instead, the dust grains sediment towards the lower layers of the photosphere, slowly depleting the upper layers. The cloud coverage is still assumed to be homogeneous. The dust grain size is determined by the equilibrium between the growth of the grains and their mixing and sedimentation. For the typical atmospheric parameters of late-L dwarfs, this results in grains with a diameter from $\sim$1 up to $\sim$10 $\mu$m. The mixing time scale is computed by extrapolating the results of hydrodynamical simulations of late-M dwarfs by \citet{2002A&A...395...99L} at lower temperatures. We can see in Figure~\ref{models} that the BT models fit quite well the overall slope of the spectrum, with only the Dusty models appearing slightly bluer than the target in the $K$-band. However, the main absorption features (i.e. the H$_2$O and CO bands, see Figure~\ref{models_K}) are poorly reproduced. The best fit parameters are very different: the Dusty models suggest $T_{\rm eff} = 1800$ K (typical of early L-dwarfs) and log(g) = 5.0, while the Settl models give $T_{\rm eff} = 1500$ K (typical of L5$-$L6 dwarfs) and log(g) = 3.5. This difference is a direct consequence of the different treatment of dust in the photosphere: the Dusty models, neglecting by assumption the dust settling, do not require low gravity to reach very red colours, while the Settl models do, to allow a greater dust content and at a higher altitude.

\subsection{A and AE models}

\citet{2011ApJ...737...34M} describe the cloud decks in the atmosphere assuming that the dust particle density follows the gas-phase pressure profile, with different ``shape functions'' to modulate the top and bottom cut-off of the clouds. In the E models, both cut-offs are very sharp; in the A models, the bottom cut-off is very sharp, while there is no top cut-off (i.e. the clouds extend all the way to the top of the atmosphere); the AE models represent an intermediate scenario, with a sharp cut-off at the bottom, and a shallow exponential cut-off at the top. The authors assumed the Deirmendjian size distribution for the dust particles \citep{2000ApJ...538..885S} with modal particle sizes of 1, 30, 60, and 100 $\mu$m. In Figures \ref{models}-\ref{models_K} we show only the A and AE models, as the E models provide a very poor fit. Both models tend to overestimate the flux in the ``blue'' portion of the spectrum (at wavelengths shorter than 1.3 $\mu$m) and also do not reproduce appropriately the shape of the $H$ and $K$-band. The strength of the CO absorption at 2.3 $\mu$m in this case is overestimated. The A model matches well the strength of the water absorption at 1.4 and 1.9 $\mu$m, while the AE model tends to overestimate the strength of both. We note that the best-fit \citet{2011ApJ...737...34M} models consider only forsterite clouds. The opacity spectrum of forsterite is very different from that of corundum and enstatite (whose extinction curves provide an excellent fit to the standard), and this could partly explain the poor quality of the fit at short wavelengths. Also, the modal grain sizes adopted vary between 1 and 100 $\mu$m, much larger than what suggested by our de-reddening. Such large grains would give a more ``grey'' extinction in the NIR range, resulting in less reddening. This could explain the very low best-fit $T_{\rm eff}$ (1200 K, typical of standard mid T-dwarfs), as this lower temperature implies an intrinsically redder spectrum, which is needed to fit the colours of ULAS~J222711$-$004547.

\subsection{Unified Cloudy Models}

In the UCM dust grains are assumed to form at the condensation temperature ($T_{\rm cond} \approx$2000 K) and then grow until they reach their critical radius (at the critical temperature $T_{\rm cr}$), at which point they precipitate  below the photosphere. The dust clouds are therefore segregated in the portion of the photosphere where $T_{\rm cr} \lesssim T \lesssim T_{\rm cond}$. In objects with higher $T_{\rm eff}$ (i.e. L dwarfs) the clouds are located higher in the photosphere, and therefore impact more the spectrum. The dust opacities are calculated assuming grains with a fixed radius of 0.01 $\mu$m, and considering only corundum, enstatite and iron. As we can see in Figure \ref{models}, the UCM fits properly the flux level in the $H$ and $K$-band, but as in the BT-Settl models, the strength of the CO absorption at 2.3 $\mu$m and of the water band at 1.4 $\mu$m is underestimated. The $H$-band appears too triangular compared to the observed one. Also, we can see in Figure~\ref{models_J} that the red-optical portion looks too smooth compared to the observed spectrum, and is slightly over-luminous. The best-fit parameters are $T_{\rm eff} = 1600$ K and log(g) = 3.5, in good agreement with the BT-Settl models. The best-fit $T_{\rm cr}$ is 1600 K, which means that the cloud deck extends all the way to the top of the photosphere.\\[\baselineskip]

In summary, all the best-fit models imply a low gravity of log(g) $\simeq$ 3.5$-$4.0 (with the exception of the BT-Dusty models), in contrast with the index-based classification as FLD-G. The temperature predicted varies significantly, from 1200 K up to 1800 K. This wide range of temperatures can be due to the different approaches to dust treatment adopted by the models considered, further highlighting the importance of condensates in the atmosphere of this red L dwarf. Another parameter that can play a role is metallicity. We must note at this point that the model grids used here are largely incomplete in the sense that they generally offer only solar metallicity spectra. An higher metallicity would facilitate the formation of dust, and could explain the remaining differences between ULAS~J222711$-$004547 and the synthetic spectra shown here. As regards the dust clouds distribution, all the best fit models imply a fully dusty photosphere, with the dust clouds extending all the way to the top of the atmosphere, although their density is assumed to be modulated in different ways, therefore resembling to our de-reddening scenario, where the dust is essentially added on the top of the dwarf's atmosphere \citep[for a more complete review and comparison of the different atmospheric models, and their dust cloud modelling, we refer the reader to][in particular their Figures 2 and 5]{2008MNRAS.391.1854H}.

\section{Conclusions}
We report the discovery of a new peculiar L dwarf, ULAS~J222711$-$004547. This object fits into the category of ``unusually red L dwarfs'', showing very red infrared colours while not displaying any particular sign of youth. Its kinematics in fact do not point towards membership of any of the known young moving groups, and its spectrum does not show the ``trademarks'' of young brown dwarfs. The index-based classification developed by \citet{2013ApJ...772...79A} suggests that the object has surface gravity consistent with the field population of L dwarfs. Current atmospheric models however suggest a low gravity nature for ULAS~J222711$-$004547, but fail to reproduce properly its spectrum. The most poorly reproduced feature are the water vapour absorption band at 1.4 $\mu$m, the ``peakiness'' of the $H$-band, and the CO absorption band at 2.3 $\mu$m. It is not excluded that a non-standard, higher-than-solar metallicity could explain these discrepancies.

However, applying a simple de-reddening curve, the spectrum of ULAS~J222711$-$004547 becomes remarkably similar to the spectra of the L7 spectroscopic standards. This result is rather surprising, as the atmospheric clouds of brown dwarfs are quite different in terms of dust grain size and density compared to the interstellar dust. Refining our de-reddening approach by using extinction curves for the most abundant dust species in the atmospheres of L dwarfs (corundum, enstatite, and iron) allowed us to obtain even higher quality fits, not only for our target but for other URLs as well. It appears that the differences in NIR colours and spectra of late-L dwarfs could be almost entirely explained by assuming a higher dust content in the photospheres of the redder objects. If our simple model is taken at face value, the grain size in the atmosphere would be typically $\sim$0.5 $\mu$m. An important point to stress is that the models that better fit the spectrum and photometry of ULAS~J222711$-$004547 are characterized by fully dusty photospheres, with the clouds extending all the way to the top of the atmosphere. This is to some extent similar to our simple de-reddening, where the dust is essentially located above the atmosphere, and suggests that URLs and giant planets are characterized by thick clouds in the uppermost layers of the atmosphere, probably caused by a combination of slightly low gravity and high metallicity.

Further study of ULAS~J222711$-$004547 and other red L dwarfs will help to understand better the role of dust clouds in ultracool atmospheres, their formation and subsequent disruption, and will lead to a better understanding of the atmospheric dynamics of gas giant planets. Variability studies in particular can help to understand the cloud dynamics in the atmosphere of these peculiar objects, highlighting the role of dust in shaping the spectra of URLs. A more detailed and rigorous modeling of the reddening induced by the dust excess will be the key to understand the nature of this objects, and could potentially lead to more robust observational constraints on the grain size distribution and abundances in the atmospheres of brown dwarfs.

\section*{Acknowledgments}
We thank the referee, Philippe Delorme, for the useful comments that have significantly improved the quality of this paper.

We thank John Gizis and Jacqueline Faherty for sharing the spectra of WISEP~J004701.06$-$680352.1 and 2MASS~J035523.37+113343.7 in electronic form.

This research is based on observations collected at the European Organisation for Astronomical Research in the Southern Hemisphere, Chile program 088.C-0048 and 186.C-0756. This publication makes use of data products from the Wide-field Infrared Survey Explorer, which is a joint project of the University of California, Los Angeles, and the Jet Propulsion Laboratory/California Institute of Technology, funded by the National Aeronautics and Space Administration.

Part of this work was carried out under the Marie Curie 7th European Community Framework Programme grant n.247593  Interpretation and Parameterization of Extremely Red COOL dwarfs (IPERCOOL) International Research Staff Exchange Scheme.

This research has made use of: the SIMBAD database operated at CDS France; the Two Micron All Sky Survey which is a joint project of the University of Massachusetts and the Infrared Processing and Analysis Center/California Institute of Technology; the \textit{Wide-field Infrared Survey Explorer}, which is a joint project of the University of California, Los Angeles, and the Jet Propulsion Laboratory/California Institute of Technology, funded by the National Aeronautics and Space Administration; the SpeX Prism Spectral Libraries, maintained by Adam Burgasser at http://pono.ucsd.edu/$\sim$adam/browndwarfs/spexprism; and, the M, L, and T dwarf compendium housed at dwarfArchives.org and maintained by Chris Gelino, Davy Kirkpatrick, and Adam Burgasser.

\label{lastpage}

\end{document}